\documentclass[5p,times,authoryear]{elsarticle}

\usepackage{amsmath}
\usepackage{bbm}
\usepackage{color}
\usepackage{graphicx}
\usepackage{algorithmic}
\usepackage{array}
\usepackage{url}
\usepackage{soul}
\usepackage{booktabs}
\usepackage{gensymb}
\usepackage{multirow}
\usepackage{xcolor,colortbl}

\definecolor{Gray}{gray}{0.85}
\newcolumntype{a}{>{\columncolor{Gray}}c}
\newcolumntype{b}{>{\columncolor{white}}c}
\makeatletter
\def\ps@pprintTitle{%
 \let\@oddhead\@empty
 \let\@evenhead\@empty
 \def\@oddfoot{}%
 \let\@evenfoot\@oddfoot}
\makeatother

\usepackage[modulo]{lineno}

\usepackage{hyperref} 
\biboptions{authoryear}

\definecolor{LightCyan}{rgb}{0.88,1,1}

\usepackage[normalem]{ulem}
\newcommand{\change}[2]{#2}

\begin{document}
\begin{frontmatter}

\title{Placenta Segmentation in Ultrasound Imaging: Addressing Sources of Uncertainty and Limited Field-of-View}

\author[kcl,tum]{Veronika~A.~Zimmer}
\author[kcl]{Alberto~Gomez}
\author[kcl,cul]{Emily~Skelton}
\author[kcl]{Robert~Wright}
\author[kcl]{Gavin~Wheeler}
\author[kcl]{Shujie~Deng}
\author[kcl]{Nooshin~Ghavami}
\author[kcl]{Karen~Lloyd}
\author[kcl]{Jacqueline~Matthew}
\author[icl,fau]{Bernhard~Kainz}
\author[tum,icl]{Daniel~Rueckert}
\author[kcl]{Joseph~V.~Hajnal}
\author[kcl,tum,helm]{Julia~A.~Schnabel}

\address[kcl]{School of Biomedical Engineering and Imaging Sciences, King's College London, London, United Kingdom}
\address[tum]{Faculty of Informatics, Technical University of Munich}
\address[cul]{School of Health Sciences, City, University of London, London, United Kingdom}
\address[icl]{BioMedIA group, Imperial College London, London, United Kingdom}
\address[fau]{FAU Erlangen-N\"{u}rnberg, Germany}
\address[helm]{Helmholtz Center Munich, Germany}

\begin{abstract}

Automatic segmentation of the placenta in fetal ultrasound (US) is challenging due to the (i) high diversity of placenta appearance, (ii) the restricted quality in US resulting in highly variable reference annotations, and (iii) the limited field-of-view of US prohibiting whole placenta assessment at late gestation. In this work, we address these three challenges with a multi-task learning approach that combines the classification of placental location (e.g., anterior, posterior) and semantic placenta segmentation in a single convolutional neural network. Through  the  classification  task the  model  can  learn  from  larger  and  more  diverse datasets while improving the accuracy of the segmentation task in particular in limited training set conditions.
With this approach we investigate the variability in annotations from multiple raters and show that our automatic segmentations (Dice of 0.86 for anterior and 0.83 for posterior placentas) achieve human-level performance as compared to intra- and inter-observer variability. Lastly, our approach can deliver whole placenta segmentation using a multi-view US acquisition pipeline consisting of three stages: multi-probe image acquisition, image fusion and image segmentation. This results in high quality segmentation of larger structures such as the placenta in US with reduced image artifacts which are beyond the field-of-view of single probes.
\end{abstract}


\end{frontmatter}

\section{Introduction}
Fetal ultrasound (US) is the primary imaging modality to monitor fetal health and development. US is relatively inexpensive and widely available, portable and safe for both mother and fetus. In the UK, all expectant mothers are offered at least two US screening examinations (in the first and second trimester of pregnancy), where the fetus' anatomy and functions are assessed and compared to normal appearances. Mainly 2D US images are acquired, due to their higher resolution, wider availability and ease of acquisition and interpretation compared to 3D US.
The rate of anomaly detection in these examinations is highly variable between institutions and sonographers, and  significantly below governmental targets for some anomalies and in certain geographical locations \citep{cardrs}. The main reason for this is that US is a highly operator- and patient-dependent modality \citep{sarris2012intra} and image quality is restricted by the limited field-of-view (FoV) later in gestation, lack of contrast, and view-dependent artifacts.

In recent years, methods from artificial intelligence research, in particular data-driven deep learning approaches, have been  successfully investigated to improve fetal screening, for example by automating standard tasks such as detection of standard fetal planes \citep{baumgartner2017sononet}, estimating fetal biometrics \citep{van2018automated,budd2019confident}, and investigating the fetal heart \citep{tan2020automated} from 2D US. 
Further, 3D US (and particularly the combination of multiple 3D views) has been exploited to improve image quality of specific body parts, like the fetal head \citep{wright2019complete} and to extend the field of view \citep{Wachinger2007,Gomez2017}.
The majority of such works focuses on the fetal body, and only few works study the placenta \emph{in utero} \citep{Torrents2019a}. Placental assessment during fetal US examination is important for the identification of pathologies which may be associated with poor fetal and/or maternal outcomes \citep{Fadl2017}. The size, shape and location of the placenta in relation to maternal orientation can be evaluated qualitatively \citep{Salomon2011}, as well as the site and type of cord insertion \citep{Kelley2020}. 
\change{}{For example, it has been shown that placental volume in the first \cite{Schwartz21} and second \cite{Quant2016} trimester can be used as a predictor for small-for-gestation (SFG) age birth weight and fetal growth restriction (FGR), as placental growth restriction precedes FGR. However, this does not hold true (especially for first trimester placental volume) for late-onset FGR and preeclampsia pregnancies \cite{Higgins2016}. The authors therefore looked at placentas at late gestation.}
Pathological conditions such as placenta accreta spectrum \citep{Jauniaux2018}, or lesions including chorioangiomata \citep{Buca2020}, that are likely to require specialist clinical management, may also be visualised.

A full evaluation of the placenta using conventional 2D US, however, is considered to be infeasible beyond the first trimester because of the limited width of the US view sector \citep{Looney2018,Songsatitanon2019,Farina2016}. As a result, the placenta can only be assessed qualitatively and in segments, which relies on a vigilant operator technique to ensure thorough coverage.

Advances in placental MR imaging including microcirculation assessment \citep{Slator2018}, 3D reconstruction \citep{Torrents2019b} and automatic segmentation \citep{Shahedi2020} are helping to increase the popularity of fetal MRI as a complementary modality to US in placental evaluation \citep{Prayer2017}. One major advantage of MRI in placental imaging is the larger FoV it affords \citep{Bulas2013}. This enables clinicians to visualise the placenta as a complete structure, facilitating a more coherent and holistic evaluation, and allowing assessment in context to other fetal and maternal structures as well \citep{Miller2006}. Nevertheless, fetal MRI has its own limitations including expense, availability, acoustic noise and sensitivity to maternal and fetal movement, which can degrade image quality \citep{Alansary2016}. Thus, US currently  remains the modality of choice for placental assessment during pregnancy.
Quantitative assessment of the placenta can be enabled by capturing and segmenting the entire placenta with multiple 3D US images acquired from different views. This is however a difficult task with a number of challenges that need to be addressed.

Automatic segmentations of the placenta are necessary to allow a quantitative assessment throughout the pregnancy. 
Early works in placenta segmentation in US images have focused on the
segmentation of anterior placentas \citep{Stevenson2015,Oguz2016}. To generalize the segmentation, semi-automatic methods have been proposed in \citep{Stevenson2015,Oguz2020}. Both methods need a manual initialization to find the position of the placenta in the image. In \citep{Oguz2018}, an ensemble of methods is proposed to increase robustness. First, an initial segmentation of the placenta is predicted using a 2D slice, and then a multi-atlas label fusion algorithm is used to provide the full segmentation in 3D. 

Convolutional neural networks (CNNs) have recently become the state-of-the-art tools for accurate segmentation \citep{Wang2020}.
When a large amount of labelled training data is available, supervised CNN approaches show impressive performance in a variety of medical image segmentation tasks, including good performances for segmenting the placenta in 3D US images 
\change{\citep{Looney2018,Yang2019},}{} 
\change{\citep{Zimmer2019,Zimmer2020}}{} \change{}{\citep{Looney2018,Yang2019,Torrents2019c,Zimmer2019,Zimmer2020,Schwartz21,Looney2021}}. 
One major drawback is, however, that accurate expert pixel-level annotations are expensive and time-consuming to acquire.

\subsection{Remaining challenges in Placenta Segmentation} \label{sec:challenges}
Three main challenges have to be overcome: 
(i) High variability in placental appearance in US; (ii) Intrinsic uncertainty and variability in placenta annotations due to poor US image quality; (iii) limited FoV in US images, prohibiting whole placenta assessments at late gestation. In the following, we describe these challenges in more detail.

\begin{figure*}
    \centering
    \setlength{\tabcolsep}{10pt}
    \begin{tabular}{cc}
        \includegraphics[scale=0.34]{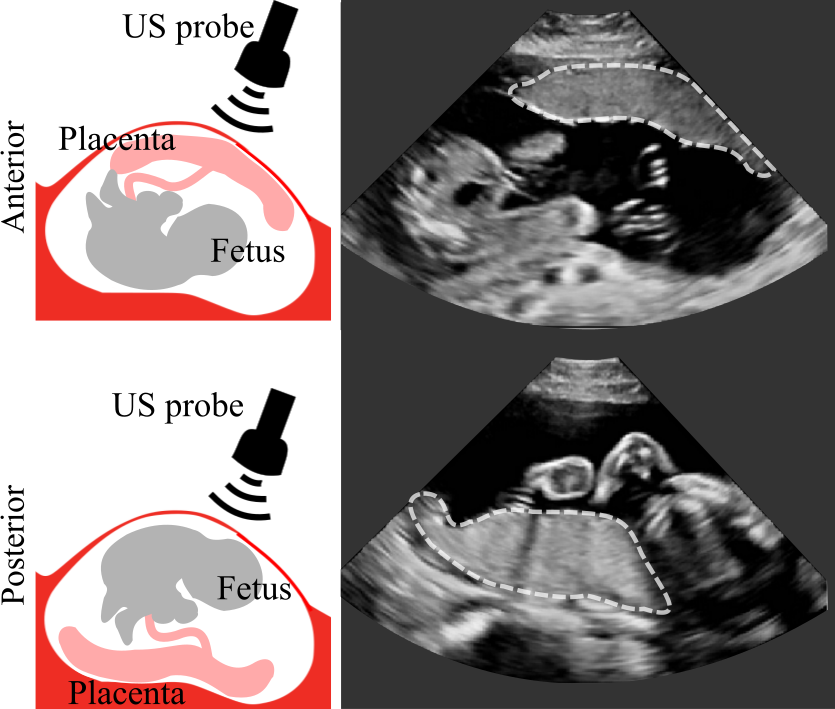} &
         \includegraphics[scale=0.34]{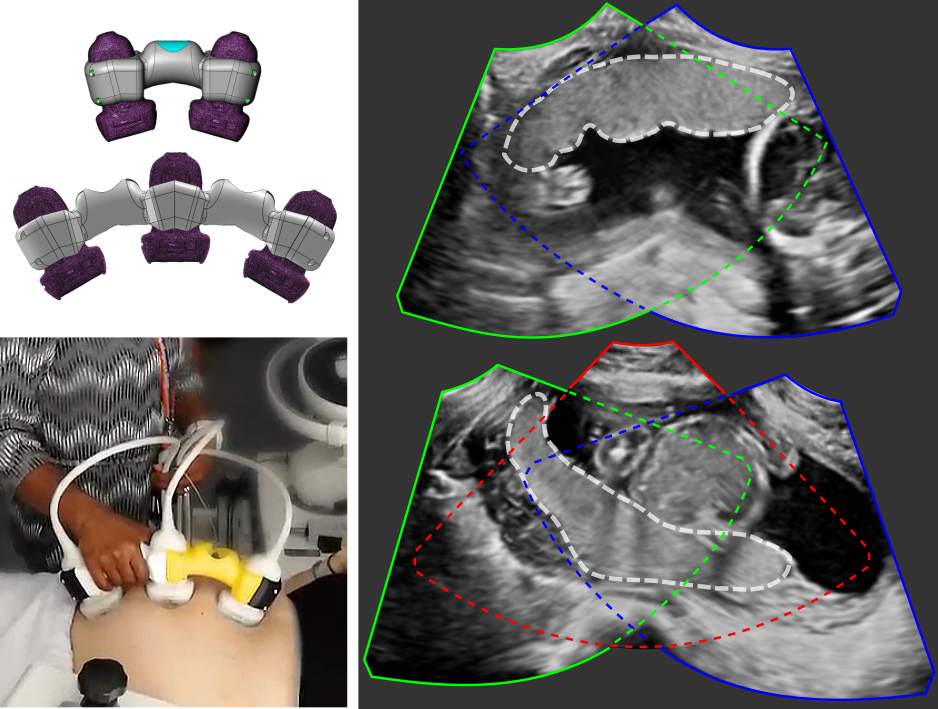} \\
         (a) Single US images & (b) Multi-view US images
    \end{tabular}
    \caption{(a): Examples of anterior (top) and posterior (bottom) placentas in ultrasound (US). (b): Design and implementation of a custom-made multi-probe holder for fetal imaging (left); two- and three-probe multi-view images (right). The placenta is delineated by white dashed lines. (All images are 3D volumes and only central 2D slices are shown.)}
    \label{fig:intro}
\end{figure*}

First, we consider variability in appearance. A major factor affecting placenta appearance in US is the location of the placenta. Anterior placentas are located at the front of the uterus towards the mother’s abdomen, and posterior placentas at the back of the uterus towards the mother’s spine, bottom) (see Fig.~\ref{fig:intro}). 
Anterior placentas are closer to the US probe, yielding higher contrast between placental and other tissues. On the other hand, the appearance of posterior placentas in US often suffers from shadows (the fetus can lie between the US probe and placenta) and attenuation artifacts. The placenta can be located in any position between the anterior or posterior of the uterine wall with the most common positions being anterior, posterior, lateral and fundal (placentas located at the left or right lateral and top of the uterus, respectively). 

Second, we consider variability and uncertainty of segmentations due to poor image quality. US images typically suffer from poor contrast, and view-dependent artifacts, which results in an intrinsic uncertainty for placenta annotation even for clinical experts. 

And third, we consider the relatively small FoV of 3D US, which normally cannot capture large structures like the second and third trimester placenta in a single image. Therefore, assessing automatically the whole placenta at late gestation is infeasible with current imaging protocols \change{}{\cite{Higgins2016}}, and it can only be assessed qualitatively in segments. 

\subsection{Related work}
Common strategies in many (medical and non-medical) applications to deal with the lack of large annotated data sets are approaches of transfer, self-supervised and multi-task learning.
In transfer learning, information and/or features can be transferred from another image domain, or another task. For the former, one starts with pre-trained models \cite{Shin2016} (e.g., pre-trained on large natural image datasets such as ImageNet) and then fine-tune the model weights on the new data. The assumption is that the pre-trained weights, even when trained on a different data domain, provide a better initialization for the optimization process during training than random weights, and that fewer data are required to achieve good performance for the final model \citep{Rajpurkar2020}.
Another approach is to use self-supervised transfer learning \citep{Shin2016,Raghu2019} to adapt the model to a new task. This involves pre-training on the target image domain, but training for a task (the pretext task) which uses different annotations that are already part of the data (or very easy to obtain). In \cite{Bai2019}, the prediction of the location of multiple anatomical positions in 2D cardiac MR images was successfully used as a pretext task to boost the accuracy of cardiac segmentation. Here, the transfer learning has been enhanced by a multi-task training strategy, where
both the pretext task and the main task are optimized together to achieve the
best performance.

In multi-task learning, the idea is to leverage knowledge and information from multiple related tasks to improve performance on all tasks \citep{Zhang2021}. The assumption is that related tasks share a common feature representation. This learning strategy is often employed, similar to transfer learning, when the data available for one or all tasks is sparse. Different to transfer learning, the knowledge between all tasks is shared and all tasks are similarly important. In medical imaging, multi-task strategies have been used successfully to detect and correct simultaneously motion-corrupted cardiac MRI sequences during reconstruction \citep{Oksuz2019}, for segmentation and bone suppression in chest X-ray images \citep{Eslami2020}, for the alignment of 3D fetal brain US images and region co-prediction \citep{Namburete2018}, for the segmentation and classification of tumors in breast US \citep{Zhou2021}, and to segment and classify CT images for COVID-19 pneumonia \citep{Amyar2020}, to just name a few.

To extend the FoV of a single image, multi-view imaging has been previously used.
In \cite{Wachinger2007,Ni2008,Gomez2017}, registration algorithm and/or tracker information were employed to align the images and provide multi-view US. The resulting image has an extended FoV, and view-dependent artifacts such as shadows can be minimized through the additional signal information from multiple views \citep{Zimmer2018}.  In \cite{wright2019complete}, many different views of the fetal head were registered to a common atlas and fused to provide a detailed, almost tomographic, image of the brain. 
Aligning US placenta remains however challenging, due to the lack of salient features to drive the registration process, and the high variability in shape, which makes it difficult, if not impossible, to define a placenta atlas space.
External tracking, on the other hand, can provide position information of the US probe but is oblivious to maternal and fetal motion.

In general, clinical adoption of segmentation methods requires that clinicians trust the segmentation results. One of the most effective ways to achieve this is by modelling the uncertainty of the estimated segmentations, and communicating this uncertainty to clinicians. Typically, two types of uncertainty are considered: (i) the \emph{aleatoric or data/intrinsic uncertainty} and (ii) \emph{epistemic or model/parameter uncertainty} \citep{Kendall2017}. The former is caused by the ambiguity and noise inherent in the data itself and is independent of the data used for training. For example, US images are often affected by artifacts and the image quality and contrast can vary greatly. The manual annotation of objects in an image might be therefore ambiguous and rather subjective. 
Also, the task of manual annotation in 3D images is difficult and their quality is dependent on annotator experience. Previous works have therefore studied the questions \emph{How good is good enough?} or \emph{How good can we actually get?} by looking at inter-rater variability \citep{Joskowicz2019}. Data uncertainty can be incorporated by multiple annotations in the training process as multiple possible labels \citep{Kohl2018} or as noisy labels \citep{Tanno2019,Zhang2020}, or estimated using test-time augmentation \citep{Wang2019}. We address the data uncertainty by exploring inter- and intra-rater variability for 3D placenta segmentation in US.
The second type of uncertainty, the model/parameter uncertainty, describes the ambiguity in the model parameters, and originates from the data used to train the model. With infinite data, the parameter uncertainty can be neglected. Bayesian approaches have been used to estimate the parameter uncertainty, such as ensemble learning \citep{Kamnitsas2017,Kohl2018} and Monte Carlo (MC) dropout as approximation to Bayesian inference \citep{Gal2016}.

\subsection{Contributions}
In this work, we propose a new method to segment 3D US images towards whole placenta segmentation in multi-view images. To achieve this, we address and overcome the three main challenges detailed in Section~\ref{sec:challenges}. 

\begin{enumerate}
    \item We address the variability in the data by leveraging the information of larger unlabeled data. 
    We propose a transfer and multi-task learning approach combining the classification of placental location and semantic placenta segmentation in a single network to capture data variability in the presence of limited training data;
    \item We explore the intra- and inter-rater variability for manual annotation of the placenta in US and study the uncertainty of automatic models. We show that the segmentations obtained by the proposed model lie within the inter-rater variability for manual placenta annotation and that the model shows less uncertainty than baseline models.
    \item We describe a multi-view US acquisition pipeline to image larger structures in US as a whole (see Fig.~\ref{fig:intro} (b)). We introduce a new US imaging technique using multiple US probes for the acquisition and fusion of multi-view images. By including the multi-task segmentation model into the multi-view imaging pipeline, we are able to extract whole placentas at late gestation.
\end{enumerate}

In particular, we propose
a multi-task approach combining the classification of placental location and semantic placenta segmentation in a single artificial neural network. The location classification as pretext task informs the network about the data variability to improve performance in unfavorable training set conditions for segmentation, which is the clinical downstream task. We discretize the placenta location in three classes: \emph{anterior}, \emph{posterior} and \emph{none}. \emph{Anterior} includes placentas located towards the front uterine wall between the fetus and the US probe, and \emph{posterior} includes placentas located towards the back uterine wall with the fetus and amniotic fluid between the placenta and the tip of the US probe  (see Fig.~\ref{fig:intro} (a)). 
\emph{None} comprises images without placental tissue, independent of the global image label from the corresponding patient.

Since the location of the placenta is typically recorded in fetal screening, training for position classification does not require any additional manual labeling. Hence many more images are available for the pretext task than for the segmentation. 

By employing this model in a multi-view US acquisition pipeline, we obtain whole placenta segmentation at late gestation, with significantly better segmentation performances than other UNet-like networks.

This study combines and expands our previous works in \cite{Zimmer2019,Zimmer2020}. In \cite{Zimmer2019}, our multi-view imaging pipeline was described for the first time. Since then, we continued to further improve on the image acquisition process, and we show here new results on a larger data set comprising multi-view images. In \cite{Zimmer2020}, we presented a first version of the multi-task model. In this work, we extended the models from 2D to 3D, added Bayesian uncertainty modelling to the UNet architecture, and extended the evaluation and discussion.

%
%

\section{Methodology}\label{sec:method}
\begin{figure*}[htb!]
    \centering
    \includegraphics[width=\linewidth]{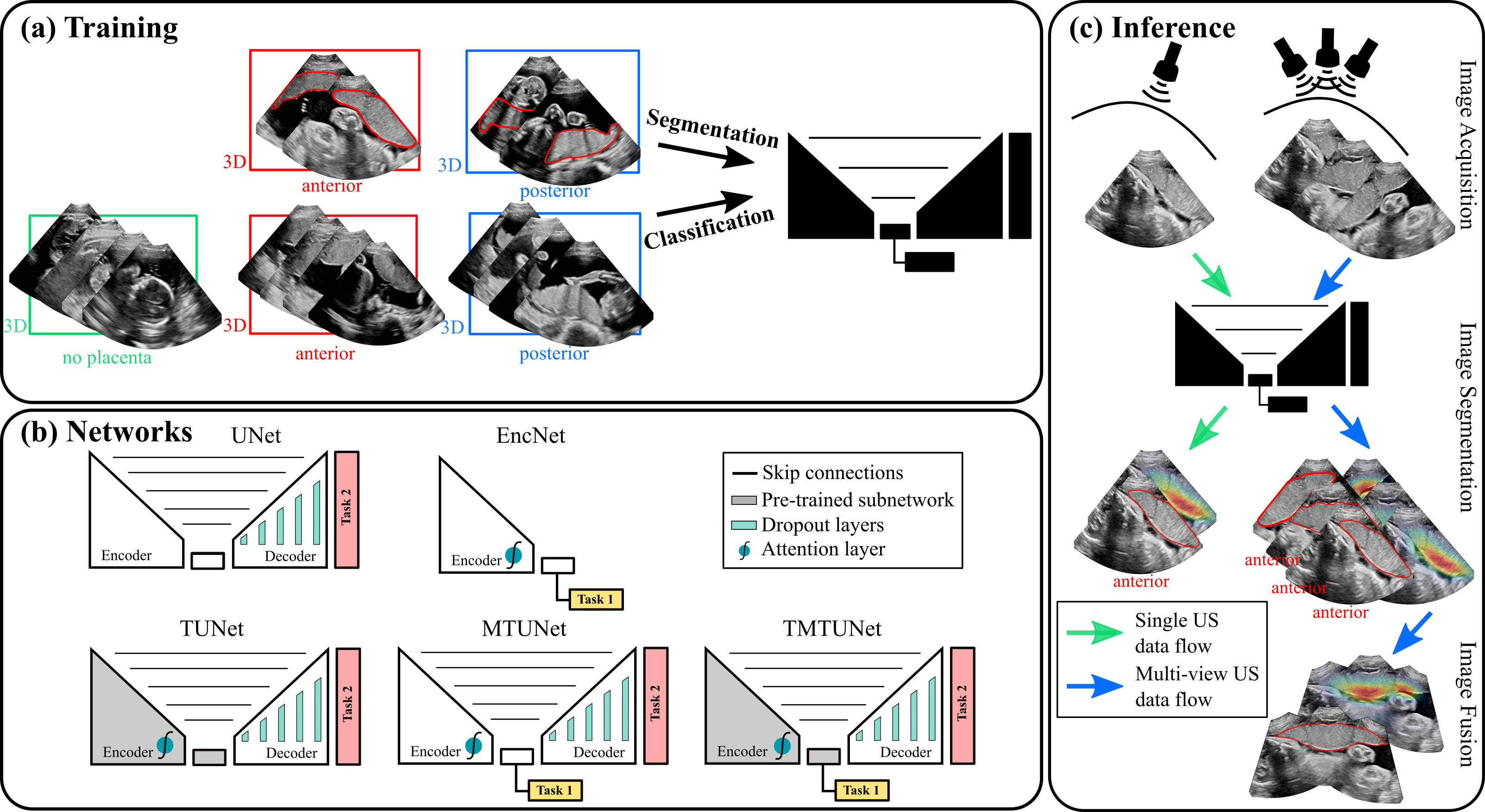}
    \caption{(a): Training data for multi-task networks using labeled datasets for segmentation and classification (classes anterior, posterior and no placenta); (b): Networks for segmentation (downstream) incorporating information from placental position classification (pretext) in different ways; (c): Inference for single and multi-view ultrasound imaging. (All images are 3D volumes, central 2D slices are shown.)}
    \label{fig:framework}
\end{figure*}

An overview of the entire image segmentation pipeline is shown in Fig.~\ref{fig:framework}. The black box represents any of the models that we compare in this paper, which are illustrated in Fig.~\ref{fig:framework} (b). The pipeline is presented in two parts. First, we describe the multi-task model to segment the placenta using positional information (Sec. \ref{sec:placenta_segmentation}), and second, we present the multi-view imaging procedure to extract the whole placenta at late gestation (Sec. \ref{sec:multiview_method}).

\subsection{Placenta segmentation and classification}
\label{sec:placenta_segmentation}

In this section we describe five CNN-based models for segmentation, classification, or both, that are evaluated and compared: \emph{UNet}, \emph{EncNet}, \emph{TUNet}, \emph{MTUNet} and \emph{TMTUNet}. These five models are illustrated schematically in Fig.~\ref{fig:framework} (b).

{\noindent\bf Notation.} 
Let us consider consider $d$-dimensional images $I_n:\Omega\subset\mathbbm{R}^d\rightarrow\mathbbm{R}$ and corresponding labels $L_n$ (here class memberships or voxel-wise segmentations). In a fully supervised strategy, the training set $\mathcal{T}=\{(I_n,L_n), n=1,\dots,N\}$ contains $N$ pairs of image and reference label, and a CNN model $f$ with parameters $\mathbf{\Theta}$ is trained to find optimal parameters $\mathbf{\Theta}^*$ to estimate for an unseen image $I$ its label $\tilde{L}=f(I,\mathbf{\Theta}^*)$. During training, a loss function $\mathcal{L}$ is optimized with respect to the parameters $\mathbf{\Theta}$. The loss function measures the agreement between reference labels $L_n$ and estimated labels $f(I_n,\mathbf{\Theta})$ over the training set $\mathcal{T}$.

{\noindent\bf Image Segmentation (\emph{UNet}).} We adapt the UNet \citep{Ronneberger2015,Ciccek2016} for our segmentation task. 
UNet has a fully convolutional encoder-decoder structure with convolutional layers, a bottleneck layer in between and skip connections from encoder to decoder. 
We use a slightly modified version where each layer consists of a residual block with strided convolutions, group normalization and ReLU activations. In the encoder, max pooling is used for downsampling. Dropout is typically used for regularization in CNNs to prevent overfitting. In training, activations of incoming features are randomly removed (with a probability of $r$). We add dropout with a dropout probability of $r=0.2$ after each layer of the decoder.

We choose as a loss function $\mathcal{L}_\text{Seg}(\tilde{S}, S)$ for training the UNet the sum of the binary cross-entropy loss and Dice loss between the output $\tilde{S}$ of the network and the manual reference segmentations $S$. This proposed model will be referred to as \emph{UNet} and will form our baseline comparison.

{\noindent\bf Image Classification (\emph{EncNet}).} \change{In image classification, labels are vectors of class membership for each image, $L_n=\mathbf{c}_n=\mathbf{e}_n\in\mathbbm{R}^C$ with the $c$-th unit vector $\mathbf{e}_c$ and $c=1,\dots,C$ denoting the class membership of image $I_n$ with $C$ classes. We denote the classification model $f_\text{Class}(I,\mathbf{\Theta}_\text{Class})$.}{In image classification, labels $L_n$ are vectors $\mathbf{c}_n\in\mathbbm{R}^C$ of class membership for each image $I_n$. They are defined as $\mathbf{c}_n=\mathbf{e}_c$ if $I_n$ belongs to class $c$ with $c=1,\dots,C$ and the $c$-th unit vector $\mathbf{e}_c$.}

Our classification (pretext) network has the same structure as the encoder of \emph{UNet} followed by a convolutional block (convolutional layer, layer normalization and ReLU), and a linear block (linear layer, layer normalization and sigmoid activation). We refer to these extra layers as the classification or pretext head. We also incorporate the attention mechanism from \cite{Jetley2018}, which not only helps in the interpretation of neural networks by providing visual clues on which image regions are important for the prediction, but also improves final classification accuracy. One attention layer (adapted to 3D volumes) was added after the third layer of the encoder. 
The trained model $f_\text{Class}(I,\mathbf{\Theta}_\text{Class})$ assigns to an unseen image $I$ two outputs: the class vector $\tilde{\mathbf{c}}$ with predicted class $\tilde{c}=\text{argmax}(\tilde{\mathbf{c}})$, and the attention map $\tilde{A}:\mathbbm{R}^d\rightarrow\mathbbm{R}$, highlighting the region in the image, which most contributed to the predicted classification. We use cross entropy as a loss function for classification, denoted by $\mathcal{L}_\text{Class}(\tilde{\mathbf{c}}, \mathbf{c})$. 
We refer to this model as \emph{EncNet}.

{\noindent\bf Learning strategies: Transfer and Multi-task Learning (\emph{TUNet}, \emph{MTUNet} and \emph{TMTUNet}).}
We explore two different strategies to incorporate the information of unlabeled data or data labeled for a different task in a supervised segmentation network: transfer and multi-task learning.

For transfer learning we use the classification of placental position ($c=0:$ anterior; $c=1:$ no placental tissue in image; $c=2:$ posterior) as a pretext task. Placental position is routinely recorded in each US scanning session and available as meta/clinical data, and does not require any additional expert labelling.
Using this strategy, the classification and segmentation tasks are trained sequentially. First, a classification network $f_\text{Class}(I,\mathbf{\Theta}_\text{Class})$ (\emph{EncNet}) is trained on the pretext task. After convergence, the encoder and bottleneck of a UNet $f_\text{Seg}(I,\cdot)$ are initialized with the optimized pretrained weights $\mathbf{\Theta}_\text{Class}^*$ and further fine-tuned on the downstream task. We refer to this model as \emph{TUNet}.

Another strategy is multi-task learning, where two or multiple tasks are optimized simultaneously. To achieve this for classification and segmentation, we added the classification pretext head after the encoder of the UNet, and added also the attention mechanism to the encoder, as shown in Fig.~\ref{fig:framework}~(b). The loss functions $\mathcal{L}_\text{Seg}(\tilde{S}, S)$ and $\mathcal{L}_\text{Class}(\tilde{\mathbf{c}}, \mathbf{c})$ are combined in a multi-task loss function $\mathcal{L}_\text{MT}(\tilde{S},\tilde{\mathbf{c}}, S, \mathbf{c})$ as
\begin{equation}
    \mathcal{L}_\text{MT}(\tilde{S},\tilde{\mathbf{c}}, S, \mathbf{c}) = \mathcal{L}_\text{Class}(\tilde{\mathbf{c}}, \mathbf{c}) + \beta \mathcal{L}_\text{Seg}(\tilde{S}, S).
    \label{eq:multitask}
\end{equation}
The parameter $\beta\in\mathbbm{R}^+$ is a weighting parameters between classification and segmentation. When $\beta>1$, it emphasizes the downstream task (placental segmentation) during training. 

The multi-task training can be combined with transfer learning by initializing the weights of the encoder and pretext head with the pretrained weights $\mathbf{\Theta}_\text{Class}^*$, and fine-tune the network using both tasks simultaneously using the multi-task loss function in Eq.~(\ref{eq:multitask}). The multi-task models are 
referred to as \emph{MTUNet} and \emph{TMTUNet} in the remainder of the paper.

\subsection{Variability and uncertainty modelling}
We adopt a simple approach towards uncertainty modelling by using dropout at test time. This will allow us to put the uncertainty of the model predictions into context with the inter- and intra-rater variability.
In standard dropout, the full activations are used at test time to obtain a single robust prediction. It is also possible \citep{Kendall2015} to use dropout at test time as an approximation to Bayesian inference. At each test run, activations are removed randomly, yielding multiple possible segmentations for the same image. These can be interpreted as MC samples obtained from the posterior distribution. In the following, we refer to this procedure during test time as MC dropout.

\subsection{Multi-view ultrasound imaging}
\label{sec:multiview_method}
Multi-view placenta imaging with US requires two steps: (i) the image acquisition using multiple probes, and (ii) the multi-view image fusion, see Fig.~\ref{fig:intro} (b) for illustration. 

{\noindent\bf Multi-probe ultrasound imaging.} 
We acquire multiple US images using an in-house US signal multiplexer which allows to connect multiple Philips X6-1 probes to a Philips EPIQ V7 US system. The multiplexer switches rapidly between up to three probes so that images from each probe are acquired in a time-interleaved fashion. 
The manual movement speeds of the transducer array is within the Nyquist sampling rates. 
Therefore, for the purpose of data processing, consecutive images are assumed to have been acquired simultaneously over a small time window. 

We designed a physical device that fixes the probes in an angle of 30$^\circ$ to each other, which ensures a large overlap between the images (see Fig.~\ref{fig:intro} (b)), and allows easy and comfortable operation. 
\change{}{\ref{sec:app:holder} with Fig.~\ref{fig:app:holder} describe and show a more detailed illustration of the probe holder design with exact measurements.}

{\noindent\bf Multi-view image fusion.}
We use a simple, but effective voxel-based weighted fusion strategy to suppress view-dependent artifacts in the images and extend the FoV. First, the images are aligned. This  can be achieved via image registration, external tracking information, or fixed multiple probes, as described in the previous section. The weight of a (transformed) data point from each single image is formulated as a function of the depth in the US image with respect to the probe position and the beam angle. In effect, image points with a strong signal (to correct for shadow artifacts) and at a position close to the center of the US frustum (where the quality of the image is typically the best) will receive higher weights. The weighted fusion method is described in detail in \cite{Zimmer2018,Zimmer2019}.
\change{}{We showed the potential of such acquired and constructed multi-view images for placental volumetry in \cite{Skelton2019}.}

%
%

\section{Materials and experiments}

\subsection{Implementation details}
We implemented the models in PyTorch 1.7.1 on a Ubuntu workstation with 48 cores of 3.80GHz and trained them on a GPU Quadro RTX 8000 48GB and CUDA 11.1. The code is publicly available\footnote{\url{https://github.com/vamzimmer/multitask_seg_placenta}}.

The hyperparameters and data augmentations for the networks were determined using the validation sets and optimized for \emph{EncNet} (for classification) and \emph{UNet} (for segmentation). We tested different numbers of layers ($\{3, 4, 5\}$) and initial feature maps ($\{4, 16, 32\}$). The best validation performance was achieved using 5 layers with (16, 32, 64, 128, 256) feature maps per layer, both for \emph{EncNet} and \emph{UNet}. For the \emph{EncNet} and the multi-task UNets, we added an attention layer in the third layer of the encoder. A dropout rate of 0.2 is used in the decoder.

The images are resampled to $128\times128\times128$. We augmented the dataset by flipping the images around the x- and z-axis (an image is not flipped upside down to keep a correct positioning of the frustum), and affine transformations (translation range of 10 voxels, rotations range of 15$^\circ$, scaling of 10 and shearing of 15 voxels).

All models are optimized using the ADAM optimizer \citep{Kingma2014} and trained until convergence. Convergence was achieved for all folds after 400 epochs (\emph{EncNet}), 100 epochs (\emph{UNet}), 50 epochs (multi-task UNets). For the classification-only \emph{EncNet}, a learning rate of $10^{-5}$ is employed, for \emph{UNet}, the initial learning rate was $10^{-4}$ and was reduced by a factor of $0.1$ at epochs 30, 70, 90, and for the multi-task and multimodel UNets, the initial learning rate was $5\cdot10^{-5}$ and was reduced by a factor of $0.1$ at epochs 20, 30, 40.

Since the number of training images for classification differs from the number of training images for segmentation, we follow the training procedure described in \cite{Bai2019}. The training alternated between the two different tasks. At each epoch, the task with the higher number of training images, here classification, was optimized for one sub-iteration and the other task, here segmentation, was optimized for $\beta$ sub-iterations. If $\beta>1$, a higher weight is assigned to the segmentation task. For our experiments, we empirically chose $\beta=4$.

The manual reference segmentations for training and evaluation were created using The Medical Imaging Interaction Tool\-kit (MITK)\footnote{\url{www.mitk.org}} \citep{Wolf2005}. 

\subsection{Data} \label{sec:data}

All data were collected as real-time 3D US image streams, on \change{subjects}{healthy volunteers} with a singleton pregnancy (at a gestational age (GA) range of 19-33 weeks). Data were collected under approved institutional ethics (NRES number 14/LO/1806) and all patients were recruited under informed consent. This study was carried out in agreement with the Declaration of Helsinki.

{\noindent\bf Datasets for classification and segmentation.}
\change{We collected images sweeping over the placenta and the fetus from 71 patients.}{We collected images from an US examination (duration 30-50 minutes) of 71 healthy volunteers. A part of the examination were sweeps covering the placenta from different directions.}
Two expert sonographers (S1 with 10 and S2 with over 15 years of experience) collected the data, and S1 and S3 (with eight years of experience) provided the manual annotations. In 35 patients an anterior placenta is observed and in 32 a posterior placenta. In four patients, 
only images without placenta visible in the FoV were used. 
For each patient, 5-30 images were selected, resulting in 1188 images in total, from which 460 show an anterior, 409 a posterior placenta, and 319 show no placental tissue.
\change{}{The images used to train and evaluate segmentation models (see below) were selected from the placental sweeps. Images which are only used in the classification task were collected from different timepoints of the examination and show very different views of the fetus and/or (unavoidable) placental tissue.}

We divided the data into two parts.
First, the whole dataset $\mathcal{I^C}$ of 1188 images with labels of the classes \emph{anterior}, \emph{posterior} and \emph{none} (no placental tissue in the image), and second, an annotated segmentation dataset $\mathcal{I^S}$ with 292 images and corresponding voxel-wise manual segmentations, manually annotated by S1 from 57 patients.  
We performed a 5-fold cross-validation where each fold divided the patients into a test, training and validation set. 
In each fold, approximately $60\%$ of the data $\mathcal{I^S}$ is used for training, and $20\%$ for both validation and testing.
Different folds had different amount of images (up to $10\%$) because of the heterogeneity of the data: each patient had a different number of images, with and without manual segmentations, and with and without placental tissue. However, we made sure that the images from individual patients were not distributed across training/validation/testing sets, the number of training images with segmentations was always the same for posterior and anterior placentas, and that each patient with manual segmentations was exactly once part of a test set.
Details about the data distribution in the folds can be found in Table~\ref{tab:app:data} in the appendix.

{\noindent\bf Multi-view data.}
A subset of the placenta sweeps described above were acquired using the multi-probe acquisition system described in Sec. \ref{sec:multiview_method}, as follows. On 21 patients, a two-probe and on 32 patients a three-probe holder was used. \change{}{ We selected 1-4 multi-view images per patient which differed in the orientation of the probes with respect to the mother's tummy.} 
This resulted in 32 two-view and 57 three-view images in total. An obstetric sonographer (S1) manually segmented the placenta in all multi-view images (50 images from anterior and 39 images from posterior placentas.)

{\noindent\bf Datasets for variability and uncertainty.}
To examine the variability and uncertainty in the segmentations, we created two additional manual reference segmentations for a subset of 53 images from 12 patients by sonographer S1 (around 1 year after the first set), and by sonographer S3, also an expert obstetric sonographer, but without prior experience with MITK. Also the multi-view images from these 12 patients were manually segmented by sonographer S1 twice. In the following, S1.1 and S1.2 denote the two sets of manual segmentations by sonographer S1.
On these additional test sets, we investigated the intra- and inter-observer variability.

\change{}{On a small subset of the multi-view data (17 two- and three-view images), a set of manual segmentations S1.2 is also created. Additionally, we created a third set of annotation (S1.3) of the same subset by fusing the manual segmentations of S1.1 from the single view images to a multi-view segmentation. 
}

\subsection{Evaluation metrics}
{\noindent\bf Segmentation and classification.}
To evaluate the segmentation performance, we use multiple criteria. To compare pairs of segmentations (an automatic and a manual (reference) segmentation), we report both the \emph{Dice} and \emph{IoU} (Intersection over Union) index as overlap measures, and the robust \emph{Hausdorff Distance} (HD) and the \emph{Average Surface Distance} (ASD) as surface metrics. The conventional HD is the maximum distance between two shapes and highly sensitive to outliers. Therefore, we report a robust HD (RHD), by considering the 95 percentile. The classification performance is assessed using the balanced accuracy, precision and F1-score.

{\noindent\bf Variability in segmentations.}
To investigate the inter-/intra-expert variability in manual segmentations, and the uncertainty in automatic segmentations, we use the \emph{Generalized Energy Distance} (GED), as described in \cite{Kohl2018}. Instead of comparing pairs of segmentations as the measures Dice, IoU, ASD and RHD, the GED compares two distributions of possible segmentations, here a set of possible automatic segmentations obtained with MC dropout and a set of manual segmentations by different annotators. It is based on a distance metric (IoU in \cite{Kohl2018} and Dice in \cite{Zhang2020}), and leverages pairwise distances. A detailed definition can be found in \cite{Kohl2018}.

To test for significance, we performed a paired Wilcoxon signed-rank test between the results of the baseline \emph{UNet} and the proposed models. We report significance at $p<0.05$ and compute the effect size $r$ as $r=|\frac{z}{\sqrt{N}}|$, where $z$ is the test statistic and $N$ is the number of paired samples. We consider the effect size as small when $r\leq0.3$, moderate when $0.3<r<0.5$ and strong when $r\geq0.5$ \citep{Cohen2013}.

\subsection{Experiments}
We perform two sets of experiments analyzing (i) classification and segmentation performance, and (ii) variability and uncertainty in manual and automatic segmentations. 

{\noindent\bf (i) Placenta classification and segmentation.} In the first set of experiments, we compare classification and segmentation performance of different variants of the models described in Sec.~\ref{sec:method}, for both individual and multi-view images.
We trained all models for segmentation (downstream task) on three different training and validation sets: set \emph{A}, set \emph{P} and set \emph{AP}. In set \emph{A}, only images with anterior placentas are used for training and validation, in set \emph{P} only images with posterior placentas, and in set \emph{AP} both types of images are used. The models are tested on both types of placentas. In the following, we use the term \emph{in-distribution} data (InD) for images whose class was part of the training set (anterior for set \emph{A} and posterior for set \emph{P}) and \emph{out-of-distribution} data (OoD) for images whose class was not part of the training set (posterior for set \emph{A} and anterior for set \emph{P}).

For classification (pretext task), the baseline \emph{EncNet} is trained on the full classification data $\mathcal{I}^C$. In the multi-task training, we restricted the number of training images for classification to avoid a large difference in numbers between the training data for the pretext and downstream tasks. Next to the 180 images with manual segmentations, we added 90 images without placental tissue and with label \emph{none} for a balanced training set for classification.

The models are tested on the complete test sets both for classification and segmentation and compared for the performance on the individual US images. As described in Sec.~\ref{sec:method} B, the resulting segmentations are then aligned and fused to obtain segmentations of the multi-view images.

{\noindent\bf (ii) Variability and uncertainty.} In a second set of experiments, we investigate the inter- and intra-rater variability of the manual segmentations and compare the variability and uncertainty in automatic segmentations. 
We measure the variability on a subset of the test data, for which three manual annotations are available, as described in Sec.~\ref{sec:data}. The intra-rater variability is the agreement between S1.1 and S1.2 and the inter-rater between S1.1 and S3. We compare the automatic segmentation to S1.1 (intra) and S3 (inter). The agreement between pairs of segmentations is measured using Dice, IoU, ASD and RHD. 

To assess the general uncertainty for placenta annotation, we compare the distributions of segmentations obtained by manual annotators and by an automatic model using $\text{GED}_\text{Dice}$ and $\text{GED}_\text{IoU}$. We compare for each training set set \emph{A}, set \emph{P} and set \emph{AP} the baseline \emph{UNet} to the best performing multi-task models. We used MC dropout during test-time to obtain a set of possible segmentations for each image.

\change{}{{\noindent\bf (iii) Downstream task: placental volume analysis.} As downstream analysis, we extract and compare placental volume from manual and automatic multi-view placenta segmentations. Additionally, we relate the volume extracted from three-probe placenta imaging with reference values throughout gestation obtained from MRI images, as reported by \cite{Leon2018}.}

\section{Results}
We first present three types of results: 1) placenta classification and segmentation when using individual images, 2) when using multi-view data, and 3) variability of the annotations and uncertainty of the segmentations.

\subsection{Placenta classification and segmentation}
\subsubsection{Individual images} $\;$

\begin{figure*}
    \centering
    \includegraphics[width=\linewidth]{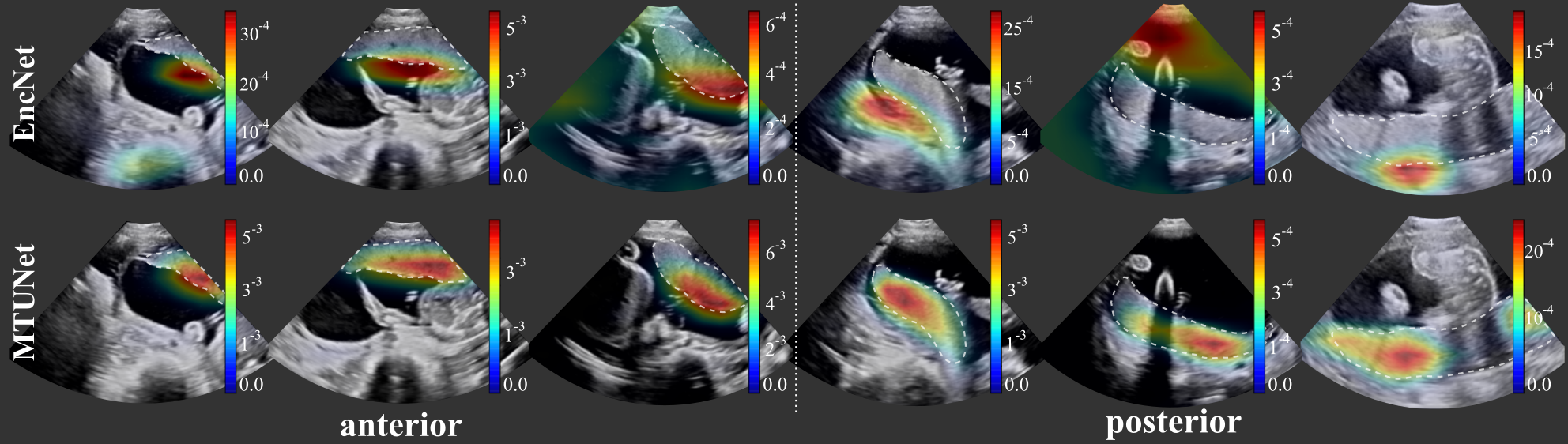}
    \caption{Attention maps \citep{Jetley2018} obtained by model \emph{EncNet} (top row) and \emph{MTUNet} (bottom row) trained on set \emph{AP} for both anterior (columns 1-3) and posterior (columns 4-6) placentas. The placenta is delineated by a white dashed line. \emph{EncNet}'s attention lies at the boundary of the placenta and surrounding tissue, \emph{MTUNet}'s on the placenta itself.  (All images are 3D volumes, central 2D slices are shown.)}
    \label{fig:attention}
\end{figure*}

{\noindent\bf Classification.}
The classification results (balanced accuracy, precision, F1-score) obtained by all models are reported in Table~\ref{tab:classification} and examples of attention maps are shown in Fig.~\ref{fig:attention}. The model \emph{EncNet} trained on the full classification training set of $817-840$ images (depending on the fold), is a strong baseline and achieved high performances on all three measures, and in particular a precision of $0.91$, $0.90$ and $0.88$ for the classes anterior, none and posterior, respectively. 

Although the training set for classification is $73.41\%$ smaller for the multi-task models, their performance on this task is competitive with the baseline \emph{EncNet} trained on the full training sets. 
Both multi-task models outperform the baseline for classes anterior and posterior, suggesting that the additional segmentation task has an influence on the performance on the pretext task. This is also confirmed by the better performance of the models trained on the segmentation set \emph{AP}. As an example, the model \emph{TMTUNet} achieved a balanced accuracy of $0.90$ for class anterior when trained on set \emph{A}, and $0.94$ when trained on set \emph{AP}. The difference between the models is that the latter uses also manual segmentations of posterior placentas during training, and this increases the performance of the classification of anterior placentas. 
A final observation is that \emph{EncNet} performs better for class \emph{none} (precision and F1-score) than the multi-task models, which can be explained by the larger number of training images, and that this class is not considered in the downstream task. 

We show attention maps obtained by models \emph{EncNet} and \emph{MTUNet} in Fig.~\ref{fig:attention}. 
In \emph{EncNet}, the model's attention lies rather at the boundary of the placenta and surrounding tissue/space than on the placenta itself. 
The additional training on segmentation in model \emph{MTUNet}, yields attention maps with good placenta localization.

\begin{table*}[t]
    \centering
    \caption{Classification performance measured by the balanced accuracy, precision and F1-score for classes anterior, none and posterior. The baseline classification model \emph{EncNet} is compared to the multi-task models trained both on classification and segmentation (\emph{MTUNet} and \emph{TMTUNet}). These models are trained on different sets for segmentation: set A (only anterior), set P (only posterior), set AP (both). Bold values indicate best performance on the corresponding class over all models. Gray boxes indicates best performance for each training set.}
    \label{tab:classification}
    \begin{tabular}{ccccc|ccc|ccc}
    \toprule
    && \multicolumn{9}{c}{Classification Performance}\\
    Train && \multicolumn{3}{c}{Balanced Accuracy} & \multicolumn{3}{c}{Precision} & \multicolumn{3}{c}{F1-score}\\
    set &Model&  anterior & none & posterior  & anterior & none & posterior & anterior & none & posterior \\
    \hline\\
    \vspace*{-0.6cm}\\
     & \emph{EncNet} & $0.91$ & $0.90$ & $\mathbf{0.93}$ & $0.91$ & $\mathbf{0.90}$ & $0.88$ & $0.91$ & $\mathbf{0.90}$ & $0.90$ \\
    \hline\\
    \vspace*{-0.6cm}\\
    A & \emph{MTUNet} & $0.83$ & $0.75$ & $0.74$ & $0.83$ & $0.75$ & $0.74$ & $0.83$ & $0.75$ & $0.74$\\
    A & \emph{TMTUNet} & \cellcolor{gray!25}$0.90$ & \cellcolor{gray!25}$0.89$ & \cellcolor{gray!25}$0.84$ & \cellcolor{gray!25}$0.91$ & \cellcolor{gray!25}$0.81$ & \cellcolor{gray!25}$0.87$ & \cellcolor{gray!25}$0.90$ & \cellcolor{gray!25}$0.84$ & \cellcolor{gray!25}$0.85$\\
    \hline\\
    \vspace*{-0.6cm}\\
    P & \emph{MTUNet} & $0.82$ & $0.77$ & $0.86$ & $0.82$ & $0.76$ & $0.86$ & $0.82$ & $0.779$ & $0.86$\\
    P & \emph{TMTUNet} & \cellcolor{gray!25}$0.92$ & \cellcolor{gray!25}$0.89$ & \cellcolor{gray!25}$0.90$ & \cellcolor{gray!25}$0.92$ & \cellcolor{gray!25}$0.87$ & \cellcolor{gray!25}$0.88$ & \cellcolor{gray!25}$0.92$ & \cellcolor{gray!25}$0.88$ & \cellcolor{gray!25}$0.89$\\
    \hline\\
    \vspace*{-0.6cm}\\
    AP & \emph{MTUNet} & $0.91$ & \cellcolor{gray!25}$\mathbf{0.93}$ & $0.90$ & $0.91$ & \cellcolor{gray!25}$0.87$ & \cellcolor{gray!25}$\mathbf{0.91}$ & $0.91$ & \cellcolor{gray!25}$0.90$ & $0.90$\\
    AP & \emph{TMTUNet} & \cellcolor{gray!25}$\mathbf{0.94}$ & $0.89$ & \cellcolor{gray!25}$0.91$ & \cellcolor{gray!25}$\mathbf{0.94}$ & $0.87$ & $0.90$ & \cellcolor{gray!25}$\mathbf{0.94}$ & $0.88$ & \cellcolor{gray!25}$\mathbf{0.91}$\\
    \bottomrule
    
    \end{tabular}
    
\end{table*}

{\noindent\bf Segmentation.}
\begin{figure*}
    \centering
    \includegraphics[width=\linewidth]{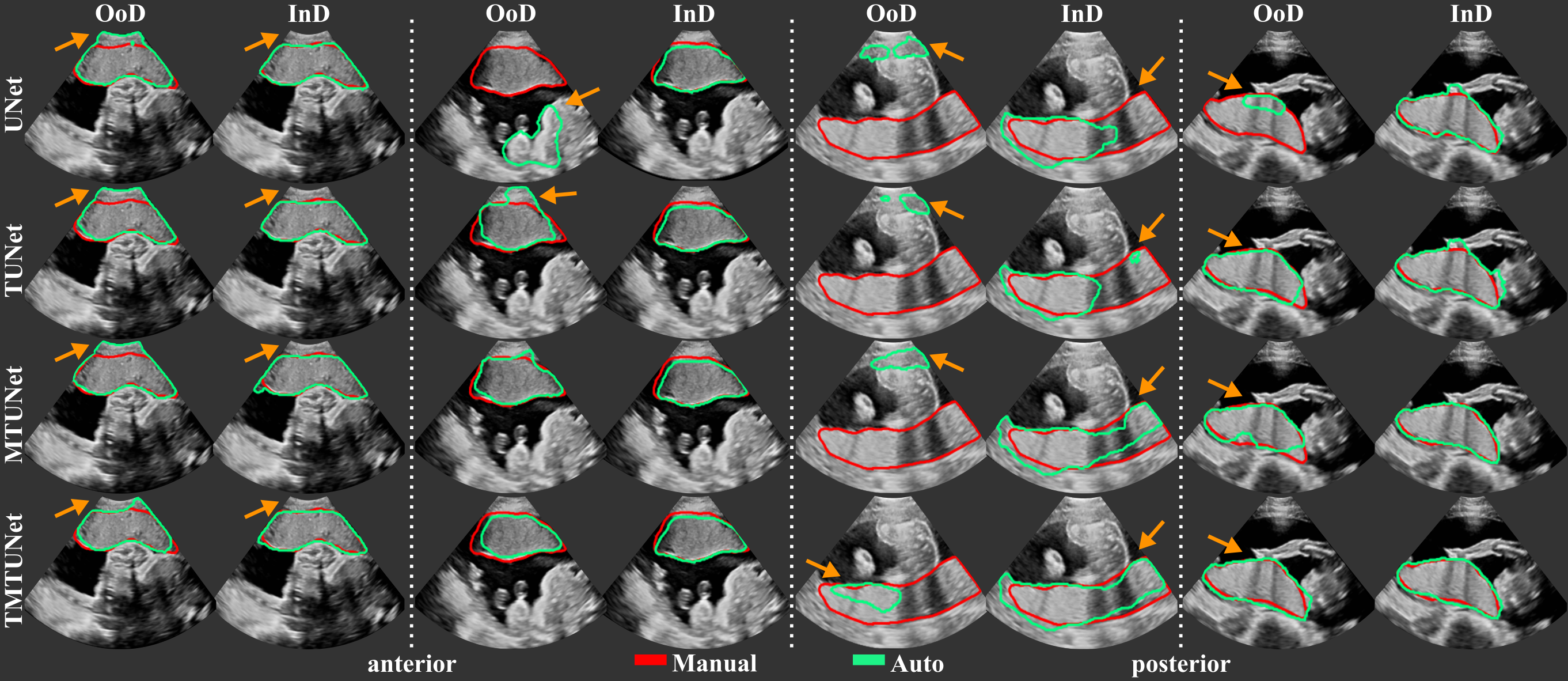}
    \caption{Examples of automatic placenta segmentations obtained by models \emph{UNet}, \emph{TUNet}, \emph{MTUNet} and \emph{TMTUNet} for pairs of \emph{in-distribution} (InD) and \emph{out-of-distribution} (OoD) test data. The orange arrows indicate areas with segmentation errors and differences between the models. (All images are 3D volumes, central 2D slices are shown.)}
    
    \label{fig:indood}
\end{figure*}
The segmentation performance of the different models measured by Dice, IoU, ASD and RHD are reported in Table~\ref{tab:segmentation} and representative segmentations comparing InD and OoD examples are shown in Fig.~\ref{fig:indood} with further examples in Fig.~\ref{fig:app:indoodex}. Results using different training and validation sets suggest that anterior and posterior placentas represent two different distributions in the data. The baseline \emph{UNet} trained on set \emph{A} (only anterior) achieves a high Dice score of $0.84$ for the InD test set (anterior), but performs poorly on the OoD set (posterior) with a Dice score of $0.26$. When trained on set \emph{P} (only posterior), the Dice score for the InD set (posterior) is $0.79$, and $0.63$ for the OoD set (anterior). 
The performance on the OoD sets is reduced with a higher standard deviation, indicating that the sets \emph{A} and \emph{P} alone are not representative enough for the segmentation of all types of placenta.
These results confirm also that it is easier to segment anterior placentas, which achieve both a higher InD and OoD Dice score. The same trend is observed for the other performance metrics (IoU, ASD, RHD) and models (\emph{TUNet}, \emph{MTUNet}, \emph{TMTUNet}).  

With the incorporation of the classification task with additional training data in models \emph{TUNet}, \emph{MTUNet} and \emph{TMTUNet}, the segmentation performances increase on the OoD data (posterior for set \emph{A} and anterior for set \emph{P}). In particular, it can be observed that with transfer learning on set \emph{A}, i.e., the initialization of the encoder weights with \emph{EncNet}, our method yields a statistically significant (moderate and strong effect size) performance increase from a Dice of $0.258$ (baseline \emph{UNet}) to $0.409$ (\emph{TUNet}) and $0.450$ (\emph{TMTUNet}). The best OoD performance are achieved with model \emph{TMTUNet}. For the InD data, the additional training data for classification, whose information is incorporated in models \emph{TUNet} and \emph{TMTUNet} via weight initialization, is not crucial and the performance increase is not statistically significant. On these data, the best performances are achieved with model \emph{MTUNet}.

When trained on set \emph{AP}, which is representative for both anterior and posterior placentas, good performances are achieved on both classes. The multi-task training improves the segmentation results, and this improvement is statistically significant for the measures Dice, IoU and ASD on all classes with model \emph{MTUNet}, the best performing model. 

Notable is that the performance of posterior placentas improve generally more with multi-task learning than the performance of anterior placentas compared to the baseline. As OoD data, posterior placentas improve the Dice score by $74.42\%$, while anterior only by $17.60\%$ with \emph{TMTUNet}. On the full set \emph{AP}, posterior improve by $2.43\%$ with \emph{MTUNet}, anterior only by $0.35\%$. 

Figure~\ref{fig:indood} visualizes examples comparing the segmentation when the images was InD or OoD data. Multi-task models, especially \emph{TMTUNet} (row 4) show a more robust performance with respect to OoD data. For example, \emph{UNet} tries to segment a posterior placenta in OoD of example 2 and an anterior placenta in OoD of example 3. Also, \emph{MTUNet} and \emph{TMTUNet}  are more robust to image artifacts, such as shadows, which is shown in InD of example 3. Further examples can be found in Fig.~\ref{fig:app:indoodex} in the appendix.

\begin{table*}[t]
\centering
\caption{Segmentation performance for \change{single and multi-view data}{\textbf{single-view data}} measured by the Dice score, Intersection-over-Union (IoU), Average Surface Distance (ASD) in mm, Robust (95\%) Hausdorff distance (RHD) in mm. The baseline (\emph{UNet}) is compared to transfer-based (\emph{TUNet}) and multi-task learning-based (\emph{MTUNet} and \emph{TMTUNet}) models. Showing performance when training on different sets: A (only anterior), P (only posterior), and AP (both)\change{, or evaluated on the multi-view data, denoted by MV}. The bold values indicate the best performance of the corresponding class over all models. Grey boxes indicate significance compared to \emph{UNet} (baseline) with a $p<0.05$ with effect sizes small, moderate ($^*$) and strong ($^{**}$).} 
\label{tab:segmentation}
\setlength{\tabcolsep}{1.5pt}
\begin{tabular}{cccc|cc|cc|cc}
\toprule
Train && \multicolumn{2}{c}{Dice} & \multicolumn{2}{c}{IoU} & \multicolumn{2}{c}{ASD (mm)} & \multicolumn{2}{c}{RHD (mm)}\\
set &Model&  anterior & posterior  & anterior & posterior & anterior & posterior & anterior & posterior \\
\hline\\
\vspace*{-0.6cm}\\
A&\emph{UNet} & $0.84\,(0.12)$ & $0.26\,(0.29)$ &  $0.74\,(0.14)$ & $0.19\,(0.23)$ & $3.09\,(7.26)$ & $33.75\,(28.03)$ &  $10.99\,(14.13)$ & $66.12\,(39.86)$  \\
A&\emph{TUNet} & $0.85\,(0.10))$ & \cellcolor{gray!25}$0.41\,(0.30)^{**}$ &  $0.75\,(0.12)$ & \cellcolor{gray!25}$0.30\,(0.25)^{**}$ & $2.69\,(3.79)$ & \cellcolor{gray!25}$24.03\,(29.33)^{*}$ &  $10.51\,(12.35)$ & \cellcolor{gray!25}$52.54\,(39.05)^{*}$  \\

A&\emph{MTUNet} & \cellcolor{gray!25}$\mathbf{0.86\,(0.09)}$ & $0.27\,(0.29)$ &  \cellcolor{gray!25}$\mathbf{0.76\,(0.12)}$ & $0.19\,(0.23)$ & \cellcolor{gray!25}$\mathbf{2.23\,(1.95)}$ & $34.31\,(29.01)$ &  \cellcolor{gray!25}$\mathbf{9.08\,(7.76)}$ & $68.24\,(39.02)$  \\
A&\emph{TMTUNet} & $0.85\,(0.11)$ & \cellcolor{gray!25}$\mathbf{0.45\,(0.29)^{**}}$ &  $0.76\,(0.12)$ & \cellcolor{gray!25}$\mathbf{0.34\,(0.26)^{**}}$ & $2.78\,(5.81)$ & \cellcolor{gray!25}$\mathbf{19.97\,(23.41)^{**}}$ &  $10.33\,(11.69)$ & \cellcolor{gray!25}$\mathbf{48.89\,(37.06)^{*}}$  \\
\hline\\
\vspace*{-0.6cm}\\

P&\emph{UNet} & $0.63\,(0.33)$ & $0.79\,(0.10)$ &  $0.52\,(0.29)$ & $0.67\,(0.12)$ & $15.44\,(22.74)$ & $4.61\,(7.10)$ &  $33.68\,(32.93)$ & $17.05\,(16.75)$  \\
P&\emph{TUNet} &\cellcolor{gray!25}$0.67\,(0.29)$ & $0.80\,(0.09)$ &  \cellcolor{gray!25}$0.56\,(0.26)$ & $0.670\,(0.12)$ & \cellcolor{gray!25}$12.25\,(19.82)$ & $3.96\,(2.46)$ &  \cellcolor{gray!25}$28.92\,(29.57)$ & $\mathbf{15.36\,(11.35)}$  \\

P&\emph{MTUNet} & \cellcolor{gray!25}$0.67\,(0.29)$ & \cellcolor{gray!25}$\mathbf{0.81\,(0.08)}$ &  \cellcolor{gray!25}$0.56\,(0.27)$ & \cellcolor{gray!25}$\mathbf{0.68\,(0.11)}$ & \cellcolor{gray!25}$12.66\,(20.62)$ & \cellcolor{gray!25}$\mathbf{3.83\,(2.46)}$ &  \cellcolor{gray!25}$29.24\,(31.10)$ & $15.77\,(13.17)$  \\

P&\emph{TMTUNet} & \cellcolor{gray!25}$\mathbf{0.74\,(0.22)^{*}}$ & $0.80\,(0.10)$ &  \cellcolor{gray!25}$\mathbf{0.62\,(0.21)^{*}}$ & \cellcolor{gray!25}$0.68\,(0.12)$ & \cellcolor{gray!25}$\mathbf{7.87\,(14.01)^{*}}$ & $4.43\,(8.26)$ &  \cellcolor{gray!25}$\mathbf{23.08\,(23.77)^{*}}$ & $15.70\,(15.24)$  \\
\hline\\
\vspace*{-0.6cm}\\

AP&\emph{UNet} & $0.864\,(0.07)$ & $0.78\,(0.12)$ &  $0.77\,(0.10)$ & $0.65\,(0.13)$ & $\mathbf{2.14\,(1.74)}$ & $4.89\,(7.26)$ &  $\mathbf{8.57\,(7.69)}$ & $18.07\,(17.33)$  \\
AP&\emph{TUNet} &$0.85\,(0.12)$ & \cellcolor{gray!25}$0.79\,(0.12)$ &  $0.76\,(0.13)$ & \cellcolor{gray!25}$0.66\,(0.13)$ & $3.07\,(8.59)$ & $4.88\,(9.10)$ &  $10.13\,(14.49)$ & $17.29\,(16.94)$  \\

AP&\emph{MTUNet} & \cellcolor{gray!25}$\mathbf{0.87\,(0.10})^{*}$ & \cellcolor{gray!25}$\mathbf{0.80\,(0.13)^{*}}$ &  \cellcolor{gray!25}$\mathbf{0.77\,(0.12)^{*}}$ & \cellcolor{gray!25}$\mathbf{0.68\,(0.14)^{*}}$ & \cellcolor{gray!25}$2.62\,(7.05)$ & \cellcolor{gray!25}$4.73\,(8.71)$ &  $9.41\,(12.06)$ & $\mathbf{16.50\,(16.84)}$  \\
AP&\emph{TMTUNet} & $0.86\,(0.10)$ & $0.79\,(0.11)$ &  $0.77\,(0.12)$ & $0.67\,(0.12)$ & $2.67\,(6.58)$ & $\mathbf{4.67\,(7.61)}$ &  $9.49\,(12.26)$ & $17.30\,(16.67)$  \\
\bottomrule
\end{tabular}
\end{table*}

\subsubsection{Multi-view images}$\;$

\begin{figure*}
    \centering
    \includegraphics[width=\linewidth]{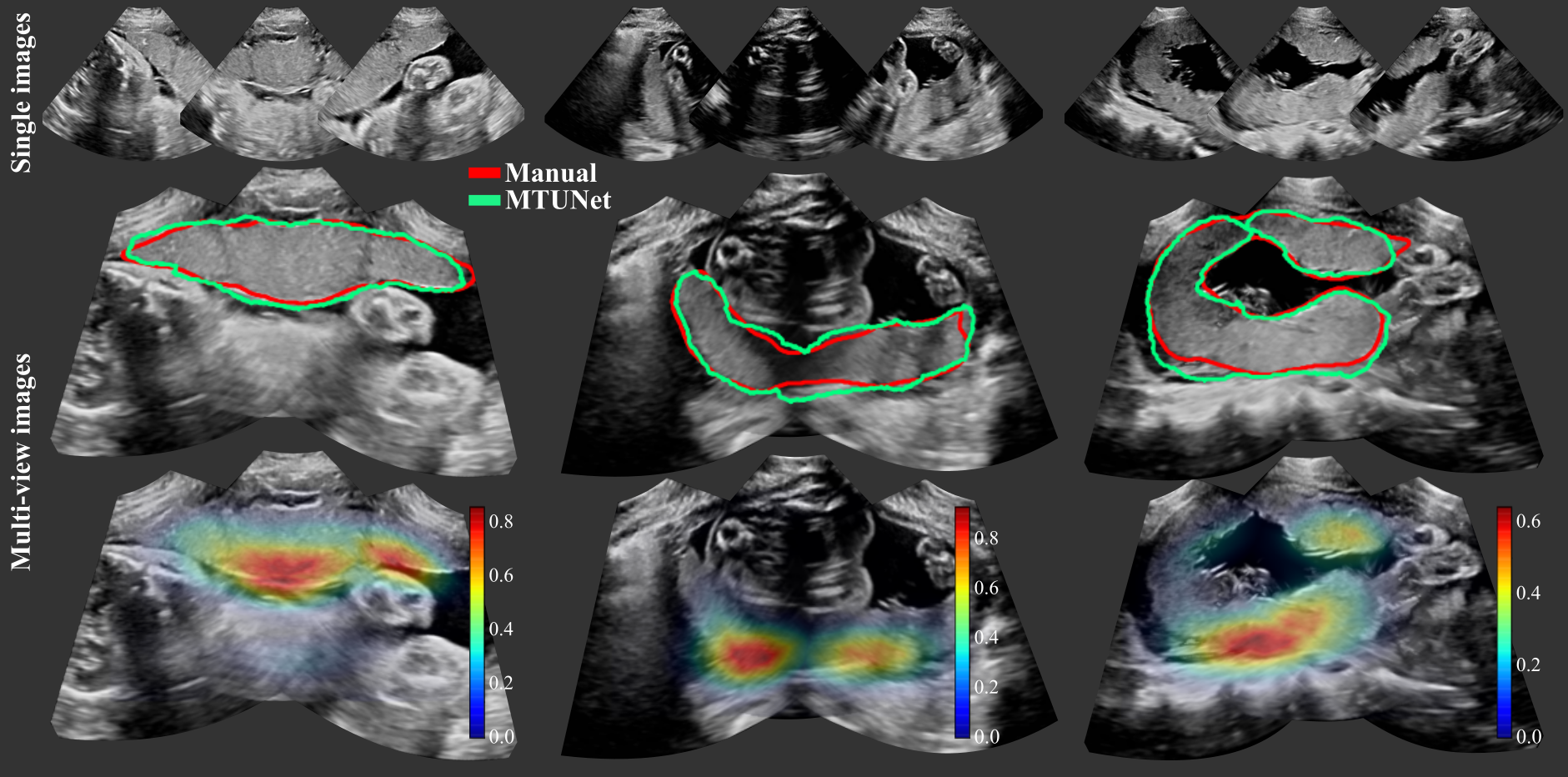}
    \caption{Three examples of multi-view images, each showing three individual images (top) and fused images with manual (in red) and automatic segmentation (model \emph{MTUNet} in green) (middle) and combined attention maps (bottom). (All images are 3D volumes, central 2D slices are shown.)}
    \label{fig:multiview}
\end{figure*}

When the spatial transformation between multiple images is known, e.g., by using a multi-probe system as described in Sec.~\ref{sec:method} for image acquisition, the segmentations in individual images can be combined to obtain the segmentation in the multi-view image. 
\change{In the following, we focus only on the models trained on the representative training set \emph{AP}.}{} The multi-view segmentation performance is reported in \change{Table~\ref{tab:segmentation} (rows starting with MV)}{Table~\ref{tab:mvsegmentation}} and representative results are shown in Fig.~\ref{fig:multiview}. 

\change{}{We observe that, in agreement with the results on single views, pre-training significantly improves the performance on OoD data, especially TMTUnet, showing a strong effect size. We would like to emphasize the performance increase on OoD data of TMTUnet trained on set P. Compared to the second best model, TUNet, the ASD is improved by 58.1\% (11.81 mm to 4.95 mm) and the RHD by 34.8\% (29.22 mm to 19.04 mm).}

\change{}{Interestingly, the performance on OoD data is in general higher on the multi-view data than on single view data. We emphasize here again that the segmentations are obtained from the single view image models and then fused for a multi-view image segmentation. The manual annotations are created on the fused images directly. We surmise that the increased performance measured on multi-view OoD data might be due to the artifact reduction in multi-view US.}

For the majority of the performance measures, the multi-task model \emph{MTUNet} performs best on both anterior and posterior placentas \change{}{on the representative training set \emph{AP}}. This is statistically significant for the measures Dice, IoU and ASD with a moderate effect size.

\begin{table*}[t]
\centering
\caption{\change{}{Segmentation performance for \textbf{multi-view data} measured by the Dice score, Intersection-over-Union (IoU), Average Surface Distance (ASD) in mm, Robust (95\%) Hausdorff distance (RHD) in mm. The baseline (\emph{UNet}) is compared to transfer-based (\emph{TUNet}) and multi-task learning-based (\emph{MTUNet} and \emph{TMTUNet}) models. Showing performance when training on different sets of single-view data: A (only anterior), P (only posterior),  AP (both) and subsequently evaluated on the multi-view data. The bold values indicate the best performance of the corresponding class over all models. Grey boxes indicate significance compared to \emph{UNet} (baseline) with a $p<0.05$ with effect sizes small, moderate ($^*$) and strong ($^{**}$).}} 
\label{tab:mvsegmentation}
\setlength{\tabcolsep}{1.5pt}
\begin{tabular}{cccc|cc|cc|cc}
\toprule
Train && \multicolumn{2}{c}{Dice} & \multicolumn{2}{c}{IoU} & \multicolumn{2}{c}{ASD (mm)} & \multicolumn{2}{c}{RHD (mm)}\\
set &Model&  anterior & posterior  & anterior & posterior & anterior & posterior & anterior & posterior \\
\hline\\
\vspace*{-0.6cm}\\
A&\emph{UNet} & $0.84\,(0.09)$ & $0.35\,(0.31)$ &  $0.74\,(0.11)$ & $0.26\,(0.25)$ & $2.88\,(2.92)$ & $26.00\,(23.94)$ &  $10.77\,(11.32)$ & $57.49\,(36.39)$  \\
A&\emph{TUNet} & \cellcolor{gray!25}$0.84\,(0.08)$ & \cellcolor{gray!25}$0.53\,(0.23)^{**}$ &  \cellcolor{gray!25}$0.73\,(0.10)$ & \cellcolor{gray!25}$0.40\,(0.21)^{**}$ & $3.12\,(2.92)$ & \cellcolor{gray!25}$13.25\,(9.38)^*$ &  $12.23\,(13.38)$ & \cellcolor{gray!25}$40.01\,(21.72)^*$  \\

A&\emph{MTUNet} & \cellcolor{gray!25}$\mathbf{0.86\,(0.07)}^*$ & $0.34\,(0.30)$ &  \cellcolor{gray!25}$\mathbf{0.75\,(0.09)}^*$ & $0.24\,(0.24)$ & $\mathbf{2.41\,(1.38)}$ & $26.96\,(23.81)$ &  $\mathbf{9.28\,(7.02)}$ & $63.00\,(38.37)$  \\
A&\emph{TMTUNet} & $0.85\,(0.08)$ & \cellcolor{gray!25}$\mathbf{0.57\,(0.23)}^{**}$ &  $0.74\,(0.10)$ & \cellcolor{gray!25}$\mathbf{0.43\,(0.23)}^{**}$ & $2.82\,(2.09)$ & \cellcolor{gray!25}$\mathbf{11.01\,(8.15)}^{**}$ &  $11.78\,(11.41)$ & \cellcolor{gray!25}$\mathbf{37.08\,(22.81)}^{**}$  \\
\hline\\
\vspace*{-0.6cm}\\

P&\emph{UNet} & $0.63\,(0.30)$ & $0.81\,(0.06)$ &  $0.52\,(0.27)$ & $0.68\,(0.09)$ & $14.94\,(21.63)$ & $4.52\,(2.92)$ &  $35.75\,(33.49)$ & $18.89\,(16.62)$  \\
P&\emph{TUNet} & \cellcolor{gray!25}$0.68\,(0.26)$ & $0.81\,(0.06)$ &  \cellcolor{gray!25}$0.56\,(0.24)$ & $0.69\,(0.09)$ & $11.81\,(19.91)$ & $4.23\,(2.54)$ &  $29.22\,(31.67)$ & $17.03\,(14.16)$  \\

P&\emph{MTUNet} & $0.64\,(0.30)$ & $0.81\,(0.06)$ &  $0.53\,(0.26)$ & $0.69\,(0.08)$ & $15.89\,(24.20)$ & $4.44\,(3.18)$ &  $36.63\,(35.87)$ & $19.05\,(18.40)$  \\
P&\emph{TMTUNet} & \cellcolor{gray!25}$\mathbf{0.77\,(0.12)}^{**}$ & \cellcolor{gray!25}$\mathbf{0.82\,(0.06)}^*$ &  \cellcolor{gray!25}$\mathbf{0.64\,(0.14)}^{**}$ & \cellcolor{gray!25}$\mathbf{0.70\,(0.08)}^*$ & \cellcolor{gray!25}$\mathbf{4.95\,(4.25)}^{**}$ & \cellcolor{gray!25}$\mathbf{3.85\,(2.35)}^*$ & \cellcolor{gray!25}$\mathbf{19.04\,(17.93)}^{**}$ & \cellcolor{gray!25}$\mathbf{15.98\,(14.72)}$  \\
\hline\\
\vspace*{-0.6cm}\\

AP&\emph{UNet} & $0.86\,(0.05)$ & $0.80\,(0.06)$ &  $0.75\,(0.07)$ & $0.67\,(0.07)$ & $2.45\,(1.25)$ & $4.76\,(2.75)$ &  $\mathbf{9.37\,(7.02)}$ & $20.55\,(16.76)$  \\
AP&\emph{TUNet} & $0.85\,(0.05)$ & $0.81\,(0.08)$ &  $0.75\,(0.08)$ & $0.68\,(0.10)$ & $2.49\,(1.30)$ & $4.50\,(3.20)$ &  $9.79\,(7.16)$ & $18.32\,(16.09)$  \\
AP&\emph{MTUNet} & \cellcolor{gray!25}$\mathbf{0.86\,(0.04)^*}$ & \cellcolor{gray!25}$\mathbf{0.82\,(0.07)^*}$ &  \cellcolor{gray!25}$\mathbf{0.76\,(0.07)^*}$ & \cellcolor{gray!25}$\mathbf{0.70\,(0.10)^*}$ & \cellcolor{gray!25}$\mathbf{2.35\,(1.42)^*}$ & \cellcolor{gray!25}$\mathbf{4.22\,(2.57)^*}$ & $9.48\,(9.46)$ & $\mathbf{17.70\,(15.01)}$  \\
AP&\emph{TMTUNet} & $0.86\,(0.05)$ & $0.80\,(0.07)$ &  $0.75\,(0.08)$ & $0.68\,(0.09)$ & $2.64\,(1.80)$ & $4.74\,(3.33)$ &  $10.97\,(11.21)$ & $19.66\,(17.35)$  \\

\bottomrule
\end{tabular}

\end{table*}

Examplary multi-view images are shown in Fig.~\ref{fig:multiview} with corresponding placenta segmentations with \emph{MTUNet} and combined attention maps.
The placenta is better visualized in the multi-view images with reduced image artifacts and an extended FoV. The multi-task model \emph{MTUNet} provides an accurate segmentation and the combined attention maps localize well the placenta. Further examples of multi-view images with corresponding segmentations can be found in Fig.~\ref{fig:app:multiview} in the appendix.

\subsection{Variability and uncertainty}

\begin{figure*}
    \centering
    \setlength{\tabcolsep}{4pt}
    \begin{tabular}{cc}
         \includegraphics[width=0.4\linewidth]{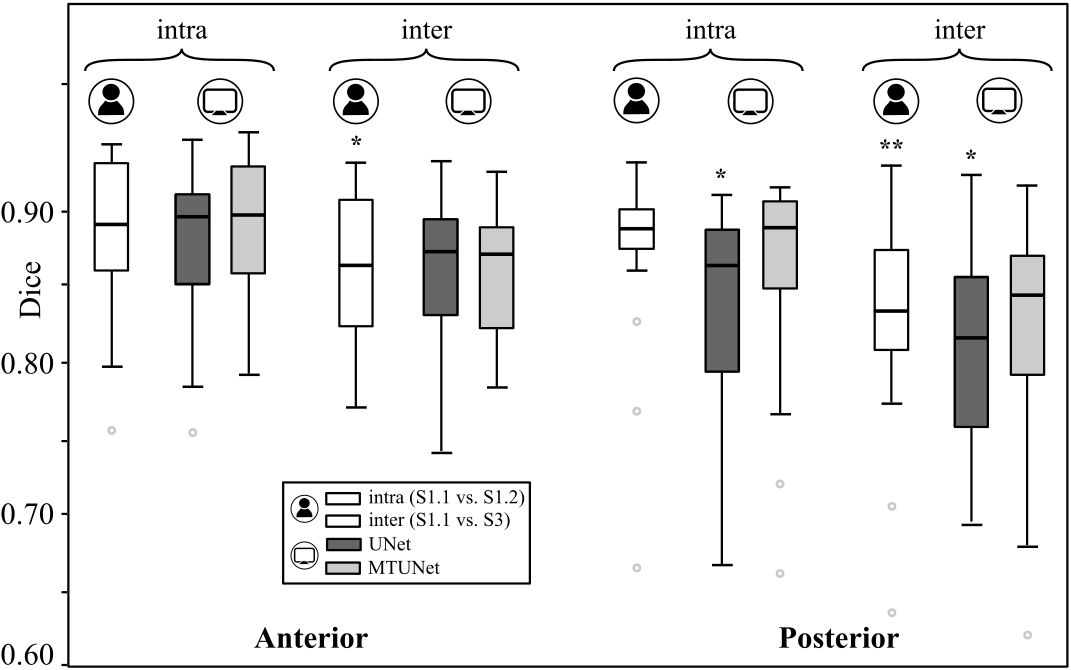}&
         \includegraphics[width=0.4\linewidth]{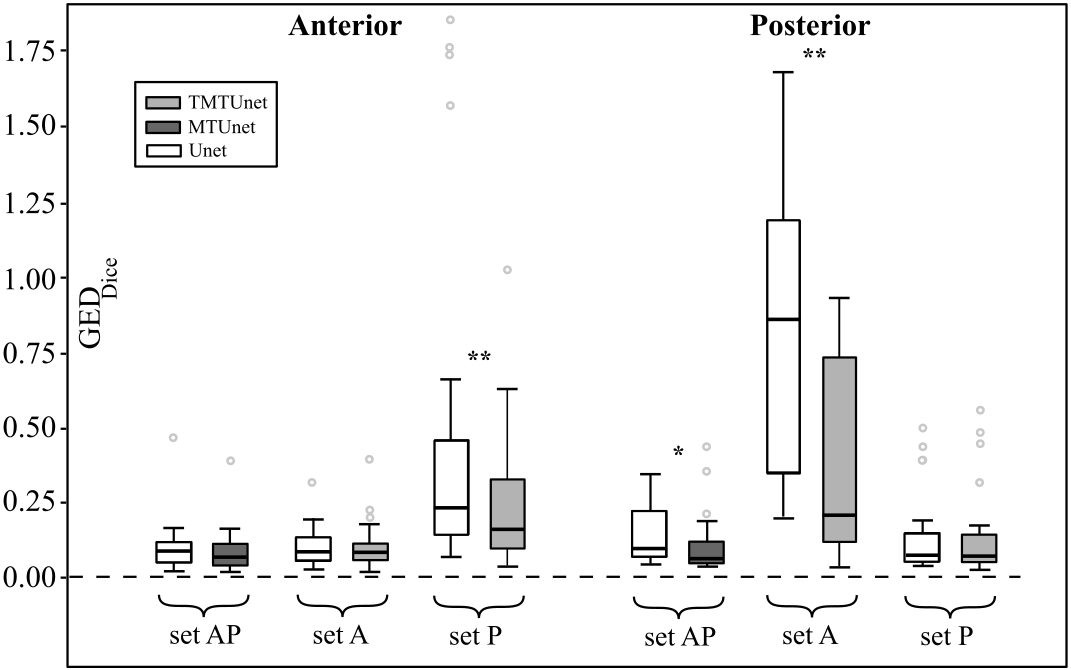}\\
         (a) Variability (manual and automatic) & (b) Uncertainty in distributions
    \end{tabular}
    \caption{(a) Variability among manual and automatic segmentations. The agreement of possible segmentation is measures using the Dice score. Manual: S1.1 vs. S1.2 (intra) and S1.1 vs. S3 (inter); \emph{UNet}/\emph{MTUNet}: S1.1 vs. \emph{UNet}/\emph{MTUNet} (intra) and S3 vs. \emph{UNet}/\emph{MTUNet} (inter). (b):  The difference in distributions between manual annotations from three raters and automatic segmentations from models \emph{UNet}, \emph{MTUNet}, and \emph{TMTUNet} with MC dropout is measured by the Generalized Energy Distance using Dice as distance measure. This is compared for models trained on sets \emph{A, P} and \emph{AP} and tested on both anterior and posterior placentas. Statistical significance between \emph{UNet} and \emph{MTUNet}/\emph{TMTUNet} is indicated by~$^*$ (moderate effect size) and $^{**}$ (strong effect size).}
    \label{fig:variabilitypw}
\end{figure*}

\begin{figure*}
    \centering
    \includegraphics[width=\linewidth]{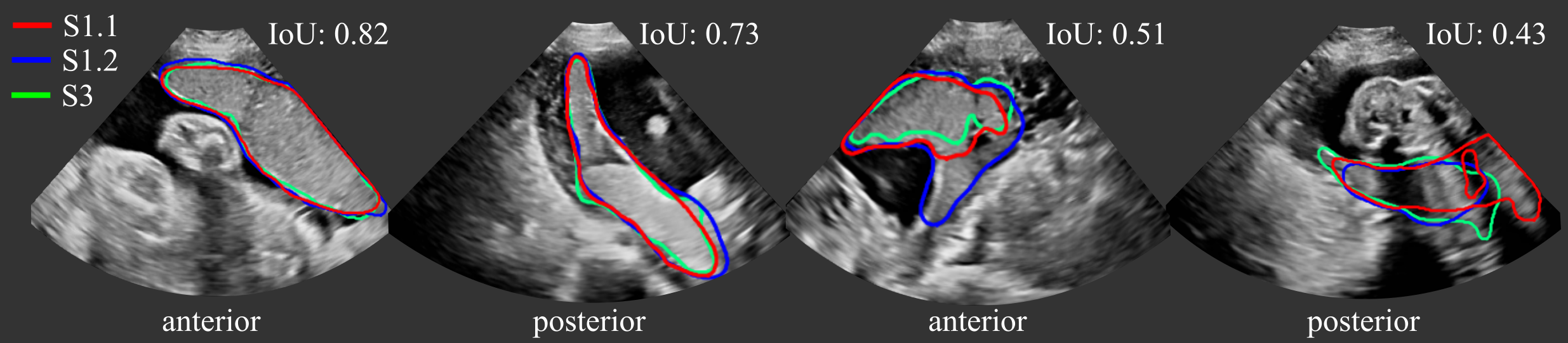}
    \caption{Manual segmentations S1.1 (red), S1.2 (blue) and S3 (green). All three segmentations agree well in (a) and (b) with an Intersection over Union (IoU) score of $0.82$ and $0.73$, respectively. Due to strong image artifacts (shadows) and/or low contrast in (c) and (d), the agreement is poorer with an IoU of $0.51$ and $0.43$. (All images are 3D volumes, central 2D slices are shown.)}
    \label{fig:manual}
\end{figure*}

We investigated the inter- and intra-observer variability for the manual annotation of placental tissue in 3D US. In each fold, we use a subset of the test set, for which three manual annotations are available, as described in Section \ref{sec:data}.
Figure~\ref{fig:variabilitypw} (a) shows the agreement of the segmentations as measured by Dice.
We compared the agreement between manual raters S1.1 and S1.2 (intra-variability) and S1.1 and S3 (inter-rater variability), and Figure \ref{fig:manual} shows examples with best and worst intra- and inter-observer agreement. 
In addition, we assess the agreement between manual and automatic segmentations (\emph{UNet} and \emph{MTUNet}), which are summerized under the term \emph{intra} with reference S1.1 and \emph{inter} with reference S3 in Fig.~\ref{fig:variabilitypw}.

Comparing the agreement between manual annotations (plain white bars in Fig.~\ref{fig:variabilitypw}), we observe that the intra-observer agreement is higher than the inter-observer agreement for all measures. The difference is statistically significant for anterior placentas with a moderate effect size and for posterior placentas with a strong effect size, denoted by one and two asterisks, respectively, above the bar for inter-rater agreement.

This suggests that the manual annotation of the placenta in US is a subjective task. In all cases and for all measures, the agreement in segmenting posterior placentas is smaller than in anterior placentas, emphasizing that the segmentation of posterior placentas is more ambiguous, possibly due to image artifacts. This is in line with the observation of the previous experiment, that the automatic segmentation models perform worse for posterior than for anterior placentas.

The intra-observer comparison of anterior placentas achieved the best agreement with a Dice of $0.89$, an IoU of $0.80$, an ASD of $1.70$ and a RHD of $12.30$. These values can be therefore interpreted as an upper bound and the range between inter- and intra-observer agreement as the desired performance of any automatic segmentation model. 
For anterior placentas, both the baseline model \emph{UNet} and our best performing model \emph{MTUNet}, as selected in the previous experiment, lie within intra- and inter-rater variability with no significant difference ($p>0.05$) between the segmentation agreements. For posterior placentas, there is a statistically significant difference (with a moderate effect size) for the baseline model \emph{UNet}, but not for \emph{MTUNet}. The multi-task approach increases the performance and reduces the variance for all measures.
The same trend is observed for IoU, ASD and RHD (see Fig.~\ref{fig:app:variabilitypw} in the appendix).

The GED scores for comparing manual and automatic segmentation distributions are shown in Fig.~\ref{fig:variabilitypw} (b).
For each training set (set \emph{A}, set \emph{P}, and set \emph{AP}) we compare the baseline \emph{UNet} to the best performing model from the first experiment (\emph{TMTUNet} for not representative training data set \emph{A} and set \emph{P}, and \emph{MTUNet} for set \emph{AP}). 

The uncertainty, as measured by GED (based on Dice as a distance measure) of the InD data, both anterior on set \emph{A} and posterior on set \emph{P} is small and comparable to the uncertainty obtained with the representative training set \emph{AP}. There is no statistical significant difference between \emph{UNet} and \emph{TMTUNet} on InD data. On OoD data, the uncertainty and variability increases and is higher for posterior than for anterior placentas. \emph{TMTUNet}, however, obtained significant lower GED scores than \emph{UNet} 
with a strong effect size both on anterior and posterior placentas. On set \emph{AP}, \emph{MTUNet} shows significantly lower GED scores for posterior placentas compare to \emph{UNet}.

The segmentation performance on this data subset is higher for all measures, classes and models comparing to the performance on the full data set as reported in Table~\ref{tab:segmentation}. This suggests that the subset contains images, showing both anterior and posterior placentas, with on average a higher image quality and less artifacts than the full data set. Thus, we surmise that our observations on variability and uncertainty would be confirmed and even stronger effects could be detected. 

\subsection{\change{}{Downstream task: placental volume analysis}}

\begin{figure*}
    \centering
    \setlength{\tabcolsep}{4pt}
    \begin{tabular}{cc}
         \includegraphics[width=0.3\linewidth]{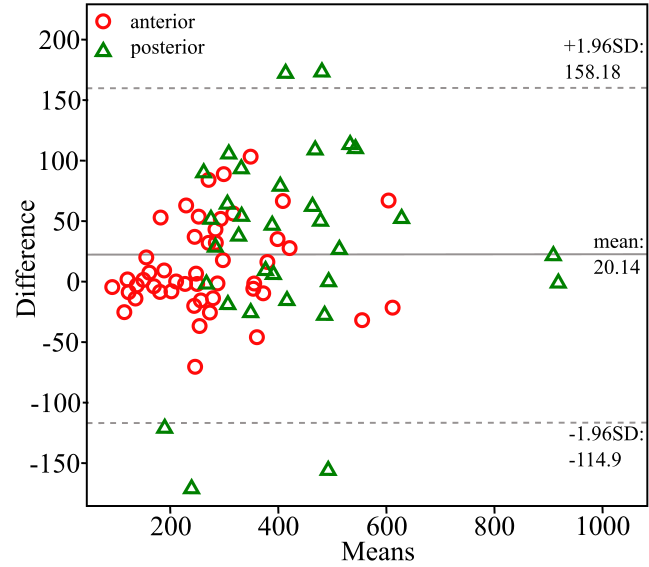}&
         \includegraphics[width=0.3\linewidth]{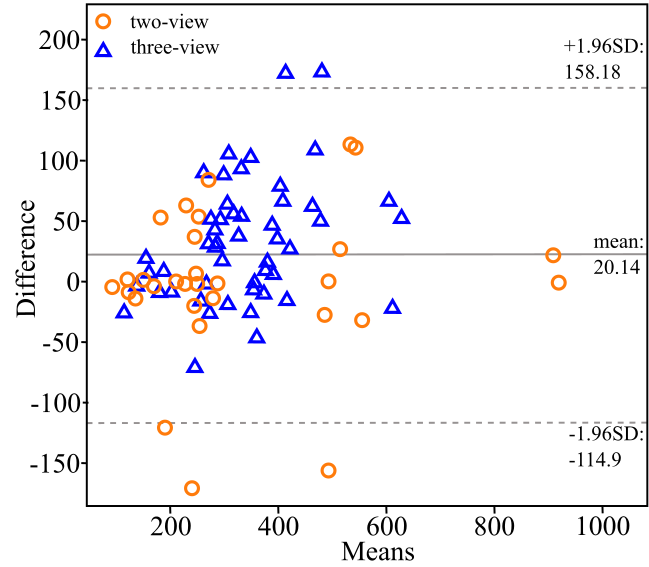}\\
         (a) S1.1 vs. MTUnet & (b) S1.1 vs. MTUnet
    \end{tabular}
    \begin{tabular}{ccc}
         \includegraphics[width=0.3\linewidth]{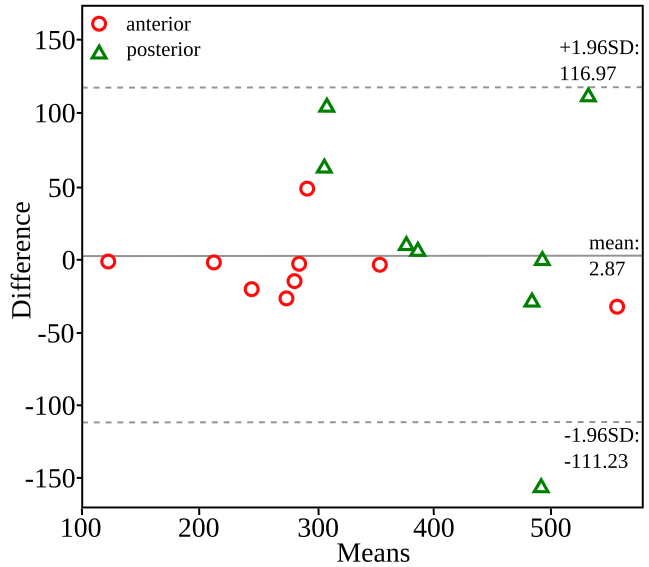}&
         \includegraphics[width=0.3\linewidth]{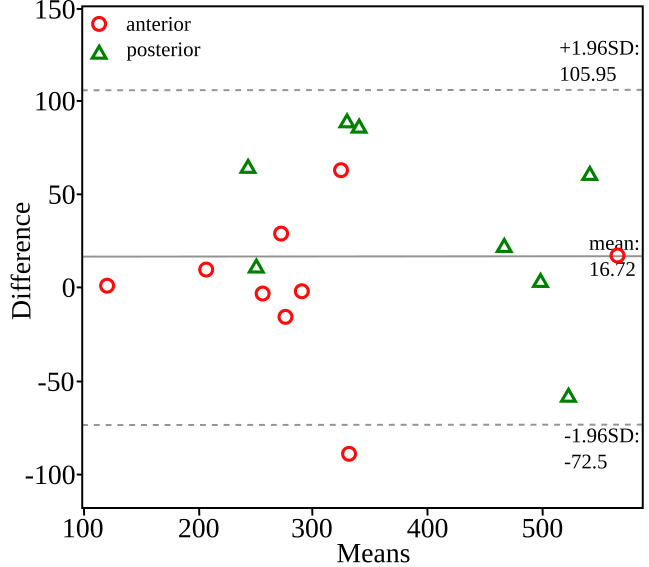}&
         \includegraphics[width=0.3\linewidth]{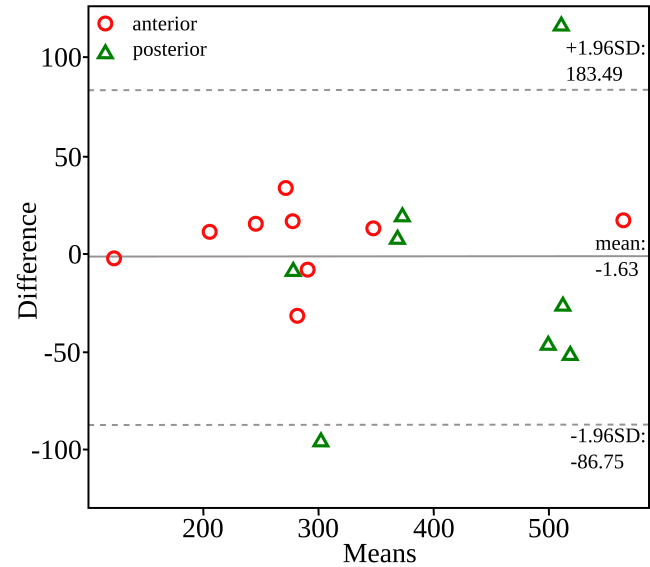}\\
         (c) S1.1 vs. MTUnet (subset) & (d) S1.1 vs. S1.2 (subset) & (e) S1.1 vs. S1.3  (subset)
    \end{tabular}
    \caption{\change{}{Bland-Altman plots comparing the placental volume (in mL) extracted from automatic and manual multi-view segmentations. (a)/(b): automatic (MTUNet trained on AP) and manual (S1.1) on the multi-view data color-coded for separating (a) anterior and posterior placentas and (b) two- and three-view images. (c)-(e): Comparison to intra-rater differences on a subset of the multi-view data. (c): automatic (MTUNet trained on AP) and manual (S1.1); (d) intra-rater (S1.1 and S1.2); (e) intra-rater (S1.1 and S1.3). S1.3 is a pseudo-manual segmentation, which is obtained by fusing the manual segmentations from S1.1 for single views.}}
    \label{fig:volume}
\end{figure*}

\change{}{Placenta segmentations can be used to extract useful clinical information, such as placental volume. In a last set of experiments, we analyze the volume computed from automatic segmentations obtained with MTUNet when trained on the representative set AP.}
\change{}{Figs.~\ref{fig:volume}~(a),(b) show Bland-Altman plots placental volume estimates obtained with MTUNet and manual segmentations S1.1 ((a) is color-coded for anterior/posterior and (b) for two-/three-view images). Outliers are mostly posterior placentas, where the image quality is reduced by artifacts. We observe that the majority of two-probe anterior placental volume estimates are relatively small. Anterior placentas are located closer to the probe (where the FoV is very narrow) and tissue is more likely missed even in two-view images.}

\change{}{We compare intra-rater variability with MTUNet in Figs.~\ref{fig:volume}~(c)-(e). The intra-rater variability is measured on a subset of the multi-view data, where two manual and one pseudo-manual annotations of rater S1 are available (S1.1, S1.2 and S1.3). For the definition of the pseudo-manual annotation see Sec.~\ref{sec:data}. We observe from the Bland-Altman plots, that the differences between MTUNet and S1.1 are comparable to the intra-rater differences (S1.1 vs. S1.2 and S1.1 vs. S1.3).
}

\begin{figure*}
    \centering
    \setlength{\tabcolsep}{4pt}
    \begin{tabular}{cc}
         \includegraphics[width=0.45\linewidth]{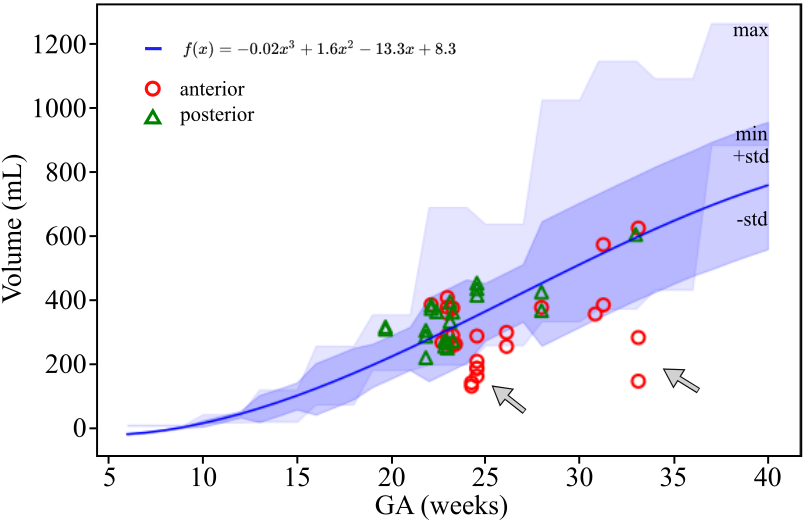}&
         \includegraphics[width=0.45\linewidth]{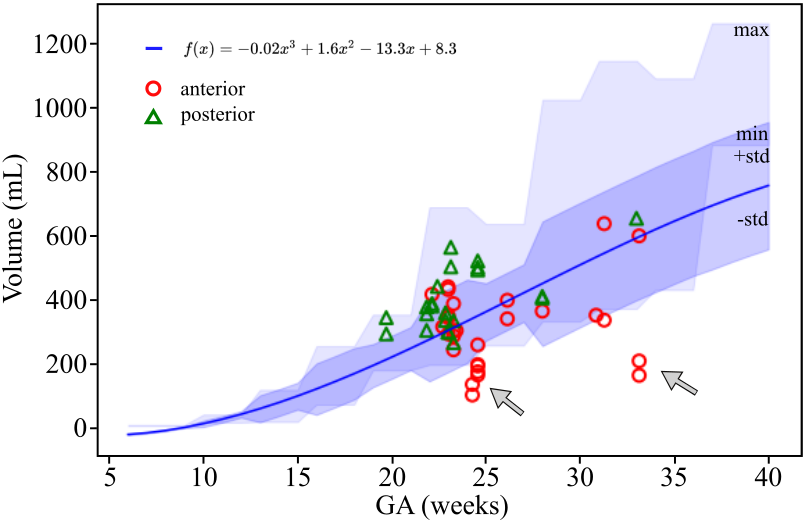}\\
         (a) Manual (S1.1) & (b) Automatic (MTUnet)
    \end{tabular}
    \caption{\change{}{Comparison of placental volumes (in mL) in multi-view (three-probe holder) images with values reported by \citep{Leon2018}. The curve $f(x)=-0.02x^3+1.6x^2-13.3x+8.3$ (blue line) was found to describe best the volume increase throughout gestation in the respective cohort. The shaded area in dark blue indicates the standard deviation and the shaded area in light blue the minimum and maximum placental volumes as reported in Table 2 in \citep{Leon2018}. (a) Manual (S1.1) and (b) automatic placenta segmentations (MTUNet trained on set \emph{AP}) show a good agreement (anterior marked as red circles and posterior as green triangles). There is also a good agreement between the volumes of the cohorts used in \citep{Leon2018} and in this study. The gray arrows indicate some outliers of anterior placentas, where some tissue is missed by the limited field-of-view close to the ultrasound probe.  }}
    \label{fig:curvelit}
\end{figure*}

\change{}{In addition, we compared the placenta volumes extracted from multi-view images (only acquired with the 3-probe holder) to values of placental volume reported in \citep{Leon2018} measured in MRI images. The authors found that the equation $f(x)=-0.02x^3+1.6x^2-13.3x+8.3$ best describes the volume increase throughout gestation in their cohort. We plot this curve with standard deviations and min/max values reported in \citep{Leon2018} together with the volumes of our cohort (three-probe holder) in Fig.~\ref{fig:curvelit} from (a) the manual annotations S1.1, and (b) the automatic segmentations of MTUNet. We observe a 
good agreement with the reference volumes S1.1 and the automatic volumes. Overall, there is a good agreement between the volumes from our cohort and the values reported in the literature. However, we observe some outliers (arrows in Fig.~\ref{fig:curvelit}) of anterior placentas. In these cases, the placenta was close to the probe, where the FoV is very narrow, and the multi-view image does not contain the whole placenta.}

\section{Discussion} 
We propose a multi-task approach combining the classification of placental position and semantic placenta segmentation in a single network. Through the classification, the model can learn from larger and more diverse datasets and improve segmentation accuracy, which are comparable to human-level performance. Our results suggest that images of anterior and posterior placentas represent two different distributions in the data. In other words they are OoD data to each other in relation to a placenta segmentation task.

We have shown that multi-task models not only improve significantly the segmentation performance on OoD data, but also the performance when trained on representative data (to a lesser extent). 
The baseline method, a \emph{UNet} trained on a large dataset including data from both distributions, 
can learn reliable segmentations. 
However, the manual voxel-wise annotation is a difficult, time-consuming and subjective task and therefore availability of such data is not always possible. 
In unfavorable training set conditions, our multi-task approach achieved up to $70\%$ improvement over the baseline. Overall, the benefits for posterior placenta segmentations were higher, as these are more affected by imaging artifacts.
To this end, our multi-task model shares the entire encoder weights for both tasks. This might not be the ideal network structure, as suggested in \cite{Guo2020}, where the authors proposed an automated method to learn the best sharing and branching configuration. This would be an interesting avenue for future work.

\begin{table*}[t]
\centering
\caption{\change{}Previous work on placenta segmentation in ultrasound with specifications about training and testing data, average performance measured by the Dice score and subjects included in the study. (CNN: convolutional neural network; RNN: recurrent neural network; cGAN: conditional generative adversarial network; CV: average performance obtained in a cross-validation strategy.)} 
\label{tab:previouswork}
\small{
\begin{tabular}{llcccll}
\toprule
\multirow{2}{*}{Reference} & \multirow{2}{*}{Method} & \multirow{2}{*}{Dice} & Training + & \multirow{2}{*}{Testing} & \multirow{2}{*}{GA} & \multirow{2}{*}{Subjects} \\
&&&Validation &&&\\
\hline\\
\vspace*{-0.6cm}\\

\cite{Stevenson2015} & Random walker, & 0.87 & - & 88 & first trimester & 3D US, singleton \\
& semi-automatic & & & & & \\
\cite{Oguz2016} & Multi-atlas label & $0.83 \pm 0.05$ & - & 14 & first trimester & 3D US, only anterior \\
& fusion & & & & & \\
\cite{Yang2019} & Multi-object, & $0.64$ & 50 + 10 & 44 & first trimester & 3D US, singleton \\
& 3D CNN + RNN & & & & (10-14 weeks) & and twin \\
\cite{Looney2018} & 3D CNN & $0.81 \pm 0.15$ & 1,097 + 100 & 1,196 & first trimester & 3D US, singleton  \\
& & & & & (11-14 weeks) & \\
\cite{Oguz2018} & 2D CNN + 3D & $0.88 \pm 0.05$ & 384 slices & 73 & first trimester & 3D US, singleton, \\
& Multi-atlas label & (anterior) & & & & 28 anterior \\
& fusion & $0.85 \pm 0.05$ & & & & 19 posterior \\
&  & (posterior) & & & & \\
\cite{Oguz2020} & semi-automatic,  & $0.82 \pm 0.06$ & - & 73 & first trimester & 3D US, singleton, \\
& Multi-atlas label & & & & (11-14 weeks) & 28 anterior\\
& fusion &   & & & & 19 posterior \\
\cite{Schwartz21} & 2D and 3D CNNs & $0.88 \pm 0.05$ & 99 & 25 & first trimester & 3D US, singleton  \\
& & & & & (11-14 weeks) & \\
\cite{Looney2021} & Single- and Multi- & $0.85 \pm 0.05$ & 1,893 + 150 & 50 & first trimester & 3D US, singleton  \\
& object, 3D CNN & & & & (11-14 weeks) &\\
\cite{Hu2019} & 2D CNN + & $0.92 \pm 0.04$ & ~954 + ~205 & ~205 & first, second and & 2D US, singleton   \\
& shadow detection & & & & trimester & and twin\\
& layer & & & & (8-34 weeks) & \\
\cite{Torrents2019c} & 3D cGAN & $0.75 \pm 0.12$ & 61 & 61 (CV) & second and third & 3D US, singleton  \\
& & & & & trimester & and twin \\
& & & & & (15-38 weeks) & \\
Ours & 3D Multi-task & $0.87 \pm 0.10$ & 1188 (292 with & 292 (CV) &  second and third & 3D US, singleton  \\
& CNN & (anterior) & segm.) &  & trimester  & \\
&  & $0.80 \pm 0.13$ &  & & (19-33 weeks) & \\
&  & (posterior) & & & & \\
\bottomrule
\end{tabular}
}
\end{table*}

\change{}{Our best performing model MTUNet achieves a Dice score of $0.87\pm0.10$ for anterior and $0.80\pm0.13$ for posterior placentas. A direct comparison to performances of other placenta segmentation models reported in the literature is difficult since they are trained and evaluated on different datasets. Table~\ref{tab:previouswork} contains a summary of previous approaches with specifications about the training and testing data, the GA of the fetus and the average Dice score achieved for placenta segmentation. The Dice scores vary from 0.64 to 0.92, and the number of data used for training and evaluation from 14 to over 1,000. The majority of other works focus on the placenta at the first trimester, and all more recent works (in the last 5 years) employ CNNs. Our segmentation results are comparable to most of these works. Note that only the work \citep{Oguz2018} separates between different positions of the placenta in the evaluation. The works \citep{Hu2019,Torrents2019c} consider both early and late gestation. The overall best performance is achieved in \citep{Hu2019} with a Dice of 0.92. However, they used 2D US (in contrast to all other methods) which has higher image quality than 3D US. In 3D US, the contrast between placenta and surrounding tissue is low, especially at early but also at late gestation. Shadow artifacts become more apparent at late gestation because of the larger size of the fetus, lying in between the US probe and the placental tissue (posterior). Also, at later gestation, only part of the placental tissue might be visible in the image (especially for our multi-view images, where the middle probe is centered on the placenta and the other two probes only “see” a small part of the placenta, which is visualized with poor contrast (as seen in Figs.~\ref{fig:multiview} and~\ref{fig:app:multiview}).}

Due to poor image quality and shadow artifacts, reproducible manual segmentation is challenging. We studied the intra- and inter-rater variability with two clinical experts. Our results show a higher inter- than intra-rater variability, more pronounced in posterior than in anterior placentas. Our proposed models lie within or very close to the manual rater agreement. When comparing distributions of segmentations, the multi-task approach yields a reduced uncertainty for OoD data than the baseline model. However, the comparison between only two different raters is rather limited and its generalizability should be investigated in the future. This could also be expanded to the fetal anatomy, where accurate segmentations are important.

We do not perform explicit uncertainty modelling or incorporate the knowledge of noisy labels into the model training, as done in \cite{Tanno2019,Zhang2020,Wang2019,Kohl2018}. To this end, we employ an approximation to Bayesian inference by using MC dropout at training and test time and interpret the variability of all possible segmentations for an image as the segmentation uncertainty. 

Multi-task models perform statistically significantly better than \emph{UNet}, however, it remains unclear if the improvement, which is rather small for the full dataset, is clinically relevant. The \emph{UNet} is a very strong baseline under ideal training set conditions. However, ideal training set conditions are hard to achieve, due to the variability of the placenta appearance in US and a multi-task approach is favoured when only limited annotated data is available.

In addition to a novel segmentation method, we describe a multi-view US acquisition pipeline consisting of three stages: multi-probe image acquisition, image fusion and image segmentation. We designed and printed new accessories for the handling of two or three probes using a standard US system. The obtained images show the anatomy from different view-directions and cover an enlarged FoV, allowing the combined imaging of larger structures in US. Using a simple but effective voxel-based weighted fusion strategy, image artifacts are reduced.

\change{}{Extracting placental volume is of clinical interest, as it is related to fetal and placental abnormalities \citep{Schwartz21,Quant2016,Higgins2016}. We conducted an analysis of placental volume extracted from manual and automatic segmentation from the multi-view images, and we showed a good agreement of these volumes with reference values extracted from MRI images \citep{Leon2018}. To this end, we have not used the segmentations/volumes to identify placenta pathologies. While the automatic detection of placental abnormalities would be the overall goal, our study only proposes a first step towards it, which is automatic placenta extraction. Our cohort consists of mainly healthy volunteers without diagnosed placental abnormalities (but blinded to fetal pathologies). A routine clinical workflow typically does not include a detailed assessment of the placenta. Our study addresses an unmet clinical need and opens up the opportunity to better study placental pathologies throughout gestation. The extension of our work to abnormal cases would be a next logical step.}

\change{}{We only included second and third trimester singleton pregnancies in our study. A next step would be to extend the models and analysis to the whole gestation by including first trimester placentas. This will in addition enable a more concise comparison to previous placenta segmentation methods. Also, it would be important to test our models on twin pregnancies. Twin pregnancies can be monochorionic (shared placenta) or dichorionic (two individual placentas). Individual placentas might pose challenges for models trained only on first trimester singleton pregnancies (when the whole placenta fits in the image). The model might not recognise a second placenta in the image. In our study, however, we use second and third trimester pregnancies. The placenta is rarely completely contained in one image and our models are trained with a variety of different views. Some contain mostly placenta, some only a small part of the placenta. Therefore, we assume that our models would also perform well for twin pregnancies, but this is speculation and has to be confirmed by future studies.}

A limitation of this study is that we consider only two separate classes: anterior and posterior placentas (next to the class \emph{none}). The placenta can be located in any position between the anterior or posterior of the uterine wall and it would be interesting to incorporate a finer classification of placentas in our models.

\section{Conclusion}
In this work we focused on US placenta imaging and address challenges arising due to the high variability of placenta appearance, the poor image quality in US resulting in noisy reference annotations, and the limited FoV of US prohibiting whole placenta assessment at late gestation. 
We propose a multi-task approach combining the classification of placental position and semantic placenta segmentation in a single network. Through the classification, the model can learn from larger and more diverse datasets and improve segmentation accuracy, which are comparable to human-level performance. Our results suggest that images of anterior and posterior placentas represent two different distributions in the data. In other words they are OoD data to each other in relation to a placenta segmentation task.

We believe that this work presents important contributions for reliable imaging and image analysis in fetal screening using US. Our proposed models show a higher robustness against poor image quality and limited data availability for training.
With accurate placenta segmentations together with a pipeline to image the whole placenta at all gestations, we enable clinicians towards a more comprehensive routine examination by considering placental health.

\section*{Acknowledgment}
This research was funded in part by the Wellcome Trust IEH Award, United Kingdom [WT 102431/Z/13/Z]. 
This work was also supported by the Wellcome/EPSRC Centre for Medical Engineering, United Kingdom [WT203148/Z/16/Z] and by the National Institute for Health Research (NIHR) Biomedical Research Centre, United Kingdom at
Guy’s and St Thomas’ NHS Foundation Trust and King’s College London, United Kingdom. The views expressed are those of the author(s) and not necessarily those of the NHS, the NIHR or the Department of Health.

\section*{References}
\bibliographystyle{plainnat}
\bibliography{biblio}

\appendix
\section{Materials and experiments}
\subsection{\change{}{Probe holder design}}\label{sec:app:holder}
\begin{figure*}
\centering
    \begin{tabular}{cc}
         \includegraphics[scale=0.14]{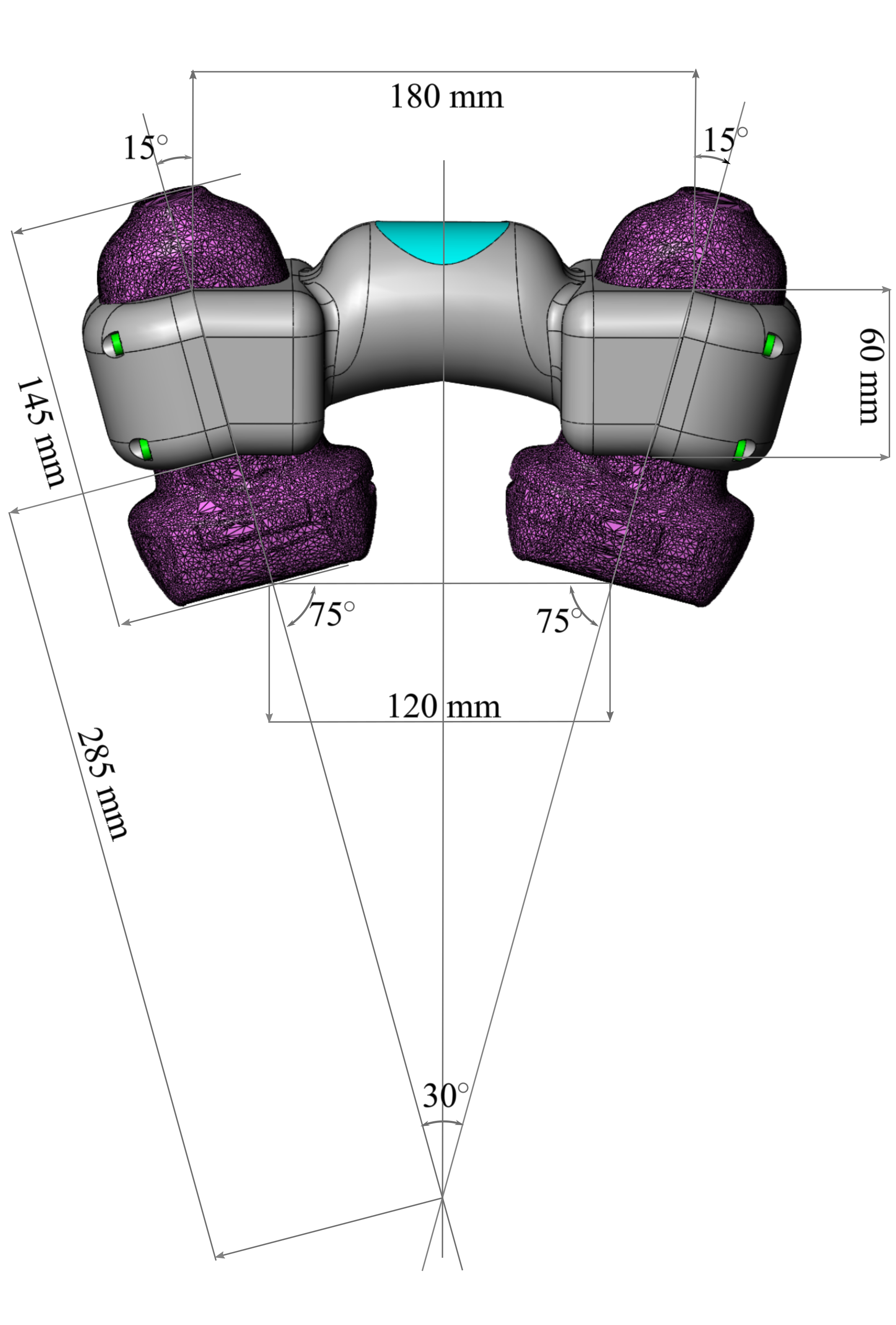} &
         \includegraphics[scale=0.14]{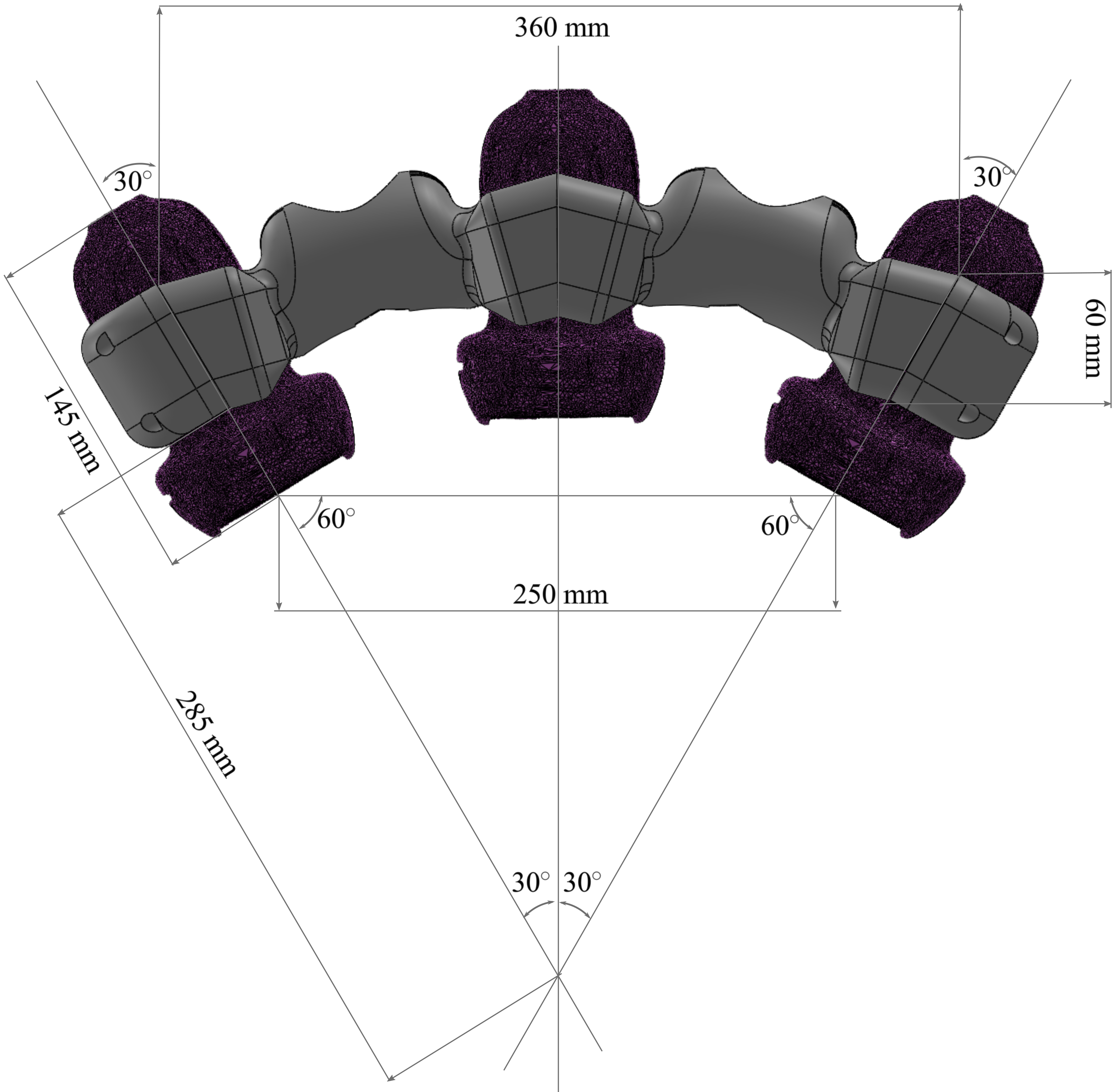}\\
         \change{}{Two-probe holder design} & \change{}{Three-probe holder design}
    \end{tabular}
    \caption{\change{}{Design of a custom-made multi-probe holder for fetal imaging.
    The probes are fixed in an angle of $30^\circ$ to each other to ensure a large overlap of the field-of-view. The system is flexible in the sense that it allows the use of two (left) or three (right) probes simultaneously.}}
    \label{fig:app:holder}
\end{figure*}

\change{}{Fig.~\ref{fig:app:holder} shows the design of the two- and three-probe holder with measurements in mm.}
\change{}{The initial design was developed on a fetal phantom in the second trimester (Kyoto Kagu Space-fan CT), and subsequently optimized with regard to comfort and usability in a clinical setting by scanning pregnant volunteers. The result is a flexible system which allows the use of two, three, or even four probes (not used in this study). We fixed the angulation between the probes so that the FoV can be extended with a known spatial alignment of the images. We chose an angle of 30$^\circ$ which empirically showed to angulate the probes sufficiently to maintain contact between the probe’s surface and maternal skin. However, other configurations are possible.}

\subsection{Data}
We perform a 5-fold cross-validation and each fold divides the patients in a test, training and validation set. 
In each fold, approximately $60\%$ of the data $\mathcal{I^S}$ is used for training, and $20\%$ for both validation and testing.
Different folds had different amount of images for validation and testing (up to $10\%$) because of the heterogeneity of the data: each patient had a different number of images, with and without manual segmentations, and with and without placental tissue. However, we made sure that the images from individual patients were not distributed across training/validation/testing sets, the number of training images with segmentations is always the same for posterior and anterior placentas, and that each patient with manual segmentations is exactly once part of a test set.

Details about the data distribution in the folds can be found in Table~\ref{tab:app:data}. 

\begin{table*}
\caption{Data splits for five folds in training, validation and testing sets for the segmentation dataset $\mathcal{I^S}$ and the classification dataset $\mathcal{I^C}$. For $\mathcal{I^S}$, the number of images are given for anterior (ant.) and posterior (post.) placentas.
For $\mathcal{I^C}$ additionally the number of images with no placental tissue visible (none) are reported.}
    \centering
    \begin{tabular}{c|cccccc|ccccccccc}
    \toprule
    & \multicolumn{6}{c}{Segmentation data $\mathcal{I^S}$}& \multicolumn{9}{c}{Classification data $\mathcal{I^C}$}\\
         &  \multicolumn{2}{c}{Training} & \multicolumn{2}{c}{Validation} & \multicolumn{2}{c}{Testing} &  \multicolumn{3}{c}{Training} & \multicolumn{3}{c}{Validation} & \multicolumn{3}{c}{Testing} \\
         & ant. & post. & ant. & post. & ant. & post. & ant. & post. & none & ant. & post. & none & ant. & post. & none\\
         \hline\\
        \vspace*{-0.6cm}\\
      Fold 1  & 90 & 90 & 34 & 24 & 30 & 22 & 286 & 290 & 241 & 101 & 64 & 49 & 89 & 55 & 29 \\
      Fold 2  & 90 & 90 & 33 & 19 & 31 & 27 & 276 & 295 & 267 & 111 & 53 & 34 & 89 & 61 & 18 \\
      Fold 3  & 90 & 90 & 34 & 21 & 30 & 25 & 285 & 298 & 240 & 102 & 51 & 48 & 89 & 60 & 31 \\
      Fold 4  & 90 & 90 & 33 & 19 & 31 & 27 & 288 & 284 & 261 & 99 & 52 & 11 & 89 & 73 & 47 \\
      Fold 5  & 90 & 90 & 32 & 23 & 32 & 23 & 287 & 296 & 267 & 98 & 55 & 19 & 91 & 58 & 33 \\
      \bottomrule
    \end{tabular}
    \label{tab:app:data}
\end{table*}

\section{Results}

\subsection{Placenta segmentation - Single images}
\begin{figure*}
    \centering
    \includegraphics[width=\linewidth]{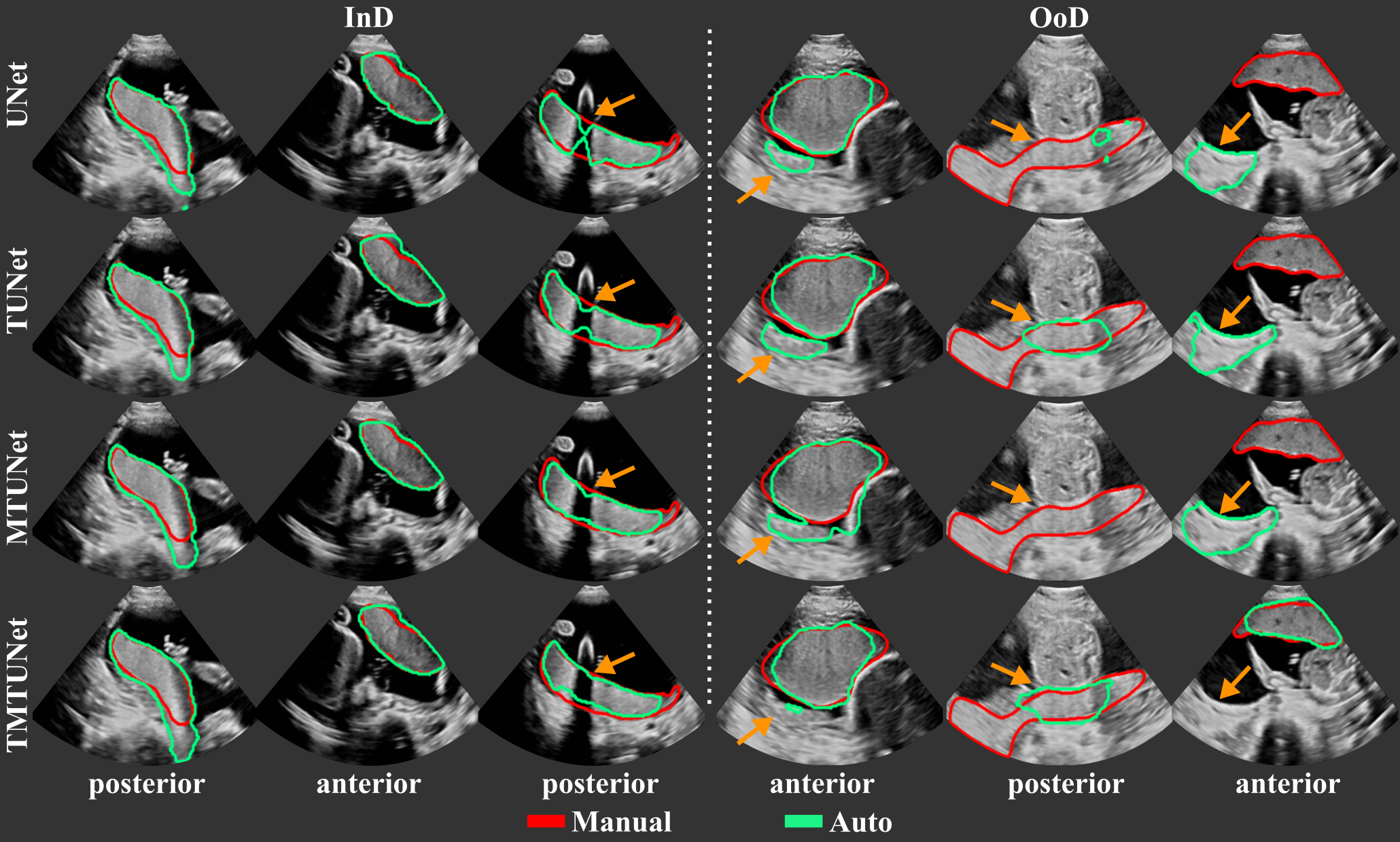}
    \caption{Examples of automatic placenta segmentations obtained by models \emph{UNet}, \emph{TUNet}, \emph{MTUNet} and \emph{TMTUNet} for \emph{in-distribution} (InD) and \emph{out-of-distribution} (OoD) test data. The orange arrows indicate areas with segmentation errors and differences between the models. (All images are 3D volumes, central 2D slices are shown.)}
    \label{fig:app:indoodex}
\end{figure*}

Figure~\ref{fig:app:indoodex} visualizes examples comparing the segmentation when the images was InD or OoD data. Multi-task models, especially \emph{TMTUNet} (row 4) show a more robust performance with respect to OoD data. Only \emph{TMTUNet} is able to localize correctly the placenta in these OoD examples. Also, \emph{MTUNet} and \emph{TMTUNet}  are more robust to image artifacts, such as shadows, which is shown in InD, last example.

\subsection{Placenta segmentation - Multi-view images}
\begin{figure*}
    \centering
	\includegraphics[width=\linewidth]{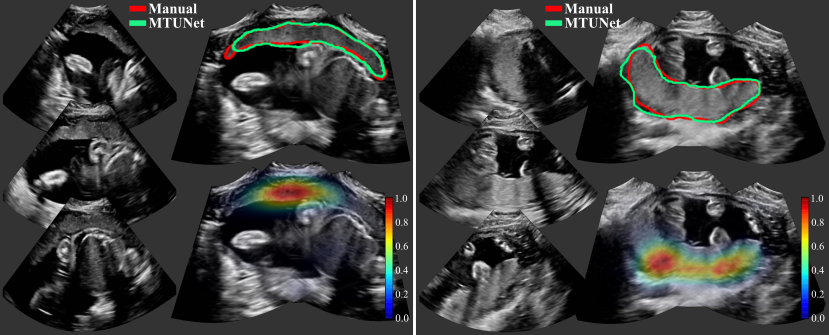}
    \caption{Four examples of multi-view images, each showing three individual images (left) and fused images with manual (in red) and automatic segmentation (model \emph{MTUNet} in green) (top right) and combined attention maps (bottom right). (All images are 3D volumes, central 2D slices are shown.)}
    \label{fig:app:multiview}
\end{figure*}

Additional examplary multi-view images are shown in Fig.~\ref{fig:app:multiview} with corresponding placenta segmentations with \emph{MTUNet} and combined attention maps.
The placenta is better visualized in the multi-view images with reduced image artifacts and an extended FoV. The multi-task model \emph{MTUNet} provides an accurate segmentation and the combined attention maps localize well the placenta.

\subsection{Variability and uncertainty}
\begin{figure*}
    \centering
    \setlength{\tabcolsep}{4pt}
    \begin{tabular}{cc}
    \includegraphics[width=0.45\linewidth]{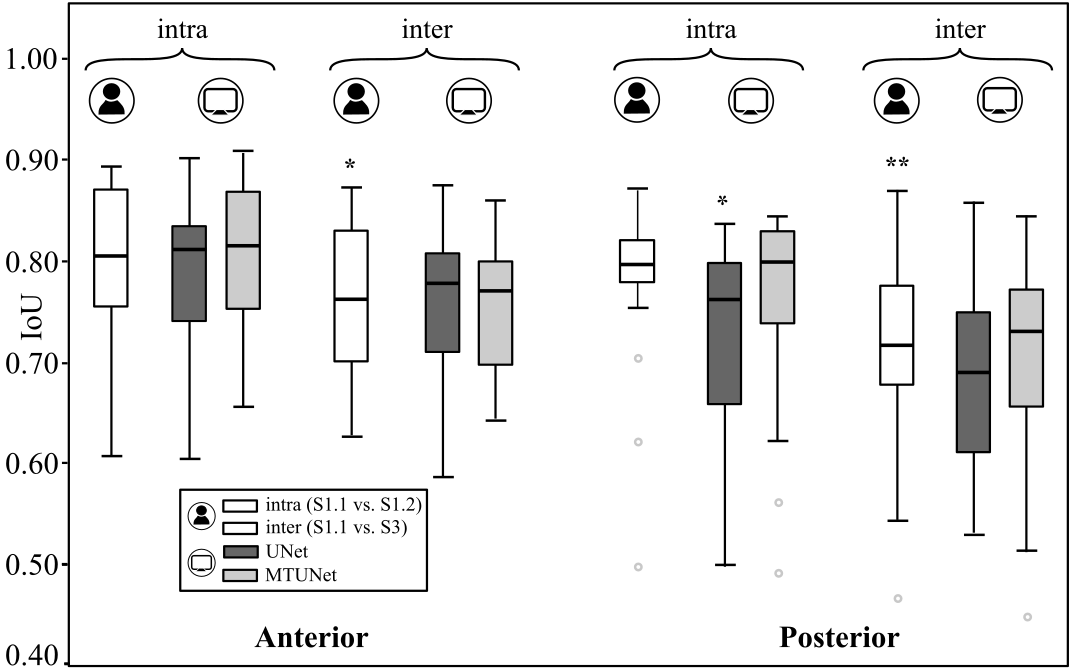}&
         \includegraphics[width=0.45\linewidth]{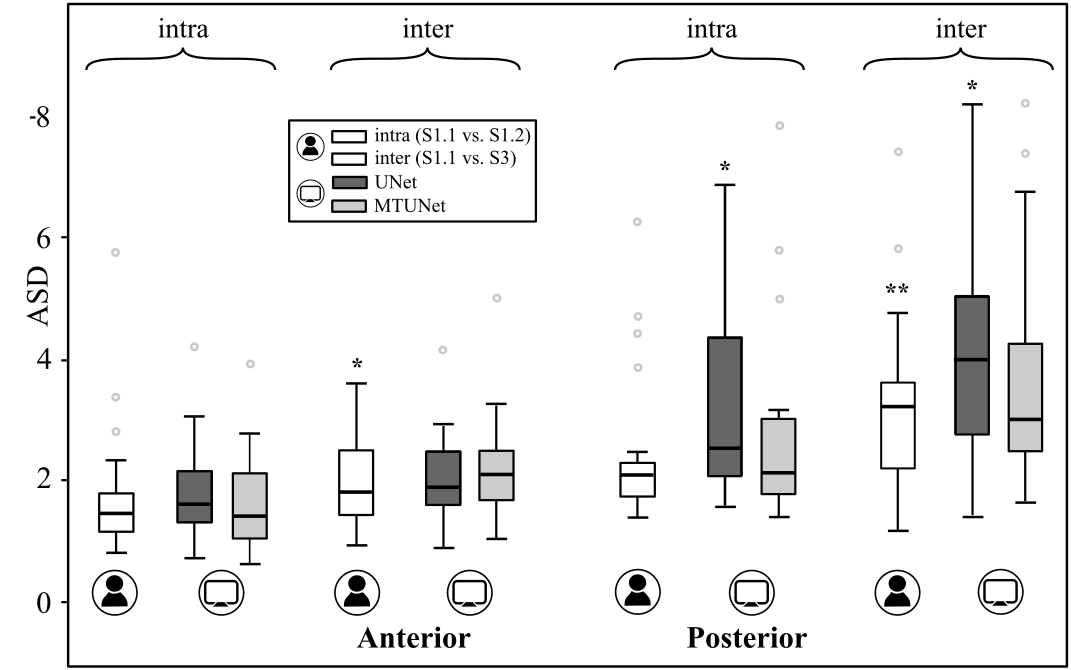}\\
         (a) IoU Variability (manual and automatic) & (b) ASD Variability (manual and automatic) \\ 
         \includegraphics[width=0.45\linewidth]{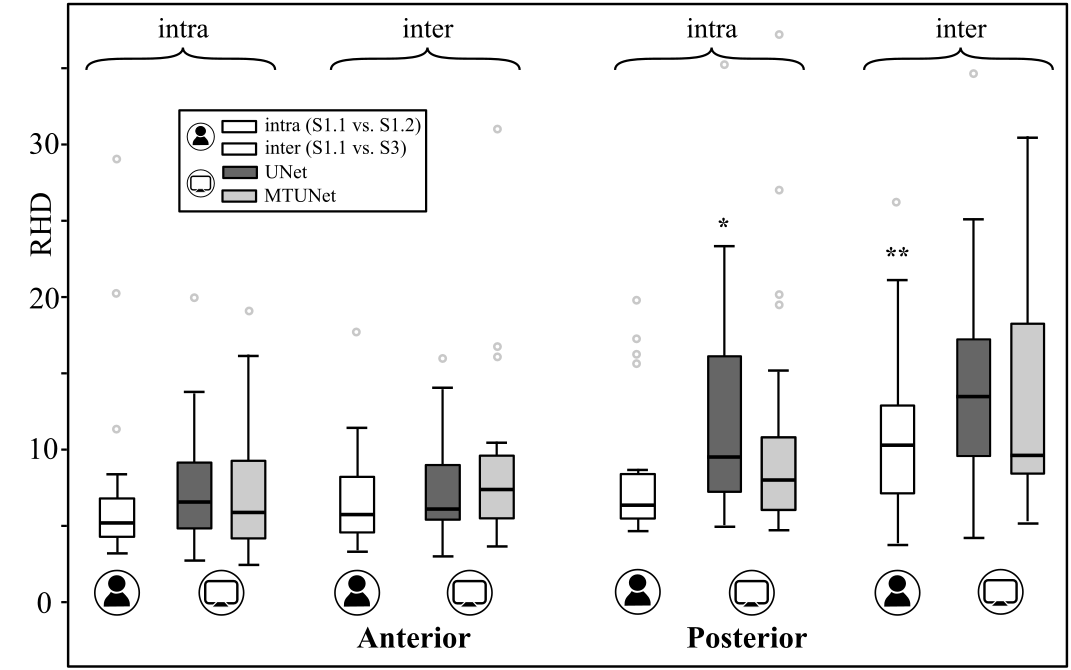} &\includegraphics[width=0.45\linewidth]{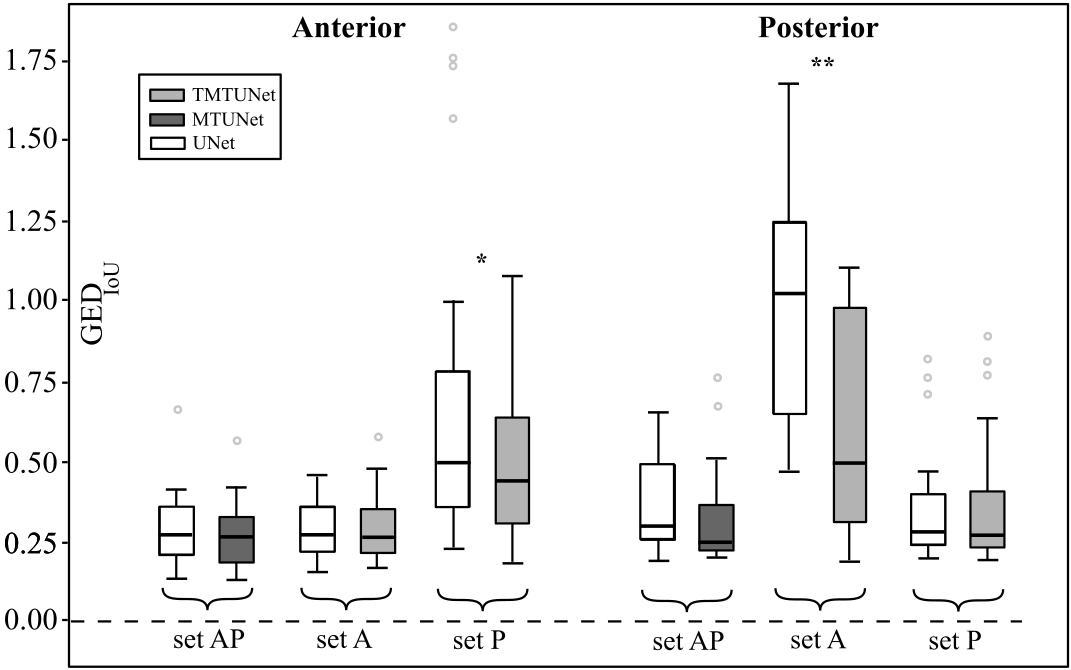}\\
         (c) RHD Variability (manual and automatic) &  (d) Uncertainty in distributions
    \end{tabular}
    \caption{(a)-(c) Variability among manual and automatic segmentations. The agreement of possible segmentation is measures using (a) the Intersection-over-Union (IoU), (b) the average surface distance (ASD) and (c) the robust Hausdorff distance (RHD). Manual: S1.1 vs. S1.2 (intra) and S1.1 vs. S3 (inter); \emph{UNet}/\emph{MTUNet}: S1.1 vs. \emph{UNet}/\emph{MTUNet} (intra) and S3 vs. \emph{UNet}/\emph{MTUNet} (inter). (b):  The difference in distributions between manual annotations from three raters and automatic segmentations from models \emph{UNet}, \emph{MTUNet}, and \emph{TMTUNet} with MC dropout is measured by the Generalized Energy Distance using IoU as distance measure. This is compared for models trained on sets \emph{A, P} and \emph{AP} and tested on both anterior and posterior placentas. Statistical significance between \emph{UNet} and \emph{MTUNet}/\emph{TMTUNet} is indicated by $^*$ (moderate effect size) and $^{**}$ (strong effect size).}
    \label{fig:app:variabilitypw}
\end{figure*}

We investigated the inter- and intra-observer variability for the manual annotation of placental tissue in 3D US. In each fold, we use a subset of the test set, for which three manual annotations are available.
Figure~\ref{fig:app:variabilitypw} (a)-(c) show the agreement of the segmentations as measured by IoU, ASD and RHD, respectively, and Fig.~\ref{fig:app:variabilitypw} (d) the difference in manual and automatic distributions (as a measure of uncertainty) measured by the Generalized Energy Distance using the Intersection-over-Union (IoU).

\end{document}